\documentclass[useAMS,usenatbib]{mn2e}

\usepackage{graphicx}
\usepackage{rotating}
\usepackage{amsmath}
\usepackage{amssymb}
\usepackage{lscape}
\usepackage{natbib}
\usepackage{keyval}
\usepackage{lastpage}
\usepackage{epstopdf}
\usepackage{array}
\usepackage{dcolumn}
\usepackage{comment}
\usepackage{pifont}
\usepackage{tikz}

\newcommand{\sfr}{\hbox{$\psi$}}
\newcommand{\mdms}{\hbox{$M_{\mathrm d}/M_\ast$}\,}
\newcommand{\mISM}{\hbox{$m_{\rm{ISM}}$}}
\newcommand{\md}{\hbox{$M_{\mathrm d}$}}
\newcommand{\mstar}{\hbox{$M_\ast$}}

\newcommand{\tick}{\ding{51}}
\newcommand{\cross}{\ding{55}}

\title[The dust budget crisis in submillimetre galaxies]{The dust budget crisis in high-redshift submillimetre galaxies}

\author[K. Rowlands et al.]
{K. Rowlands$^{2,1}$\thanks{E-mail:ker7@st-andrews.ac.uk}, 
 H.~L. Gomez$^{3}$, ~L. Dunne$^{4}$, A. Arag\'{o}n-Salamanca$^{2}$, S. Dye$^{2}$, \newauthor S. Maddox$^{4}$, E. da Cunha$^{5}$, P. van der Werf$^{6}$
\\
   $^{1}$(SUPA) School of Physics \& Astronomy, University of St Andrews, North Haugh, St Andrews, KY16 9SS, UK \\
   $^{2}$School of Physics \&\ Astronomy, The University of Nottingham, University Park Campus, Nottingham, NG7 2RD, UK \\ 
   $^{3}$School of Physics \&\ Astronomy, Cardiff University, Queens Buildings, The Parade, Cardiff, CF24 3AA, UK \\
   $^{4}$Department of Physics and Astronomy, University of Canterbury, Private Bag 4800, Christchurch, New Zealand \\
   $^{5}$Max Planck Institute for Astronomy, Konigstuhl 17, 69117, Heidelberg, Germany \\
   $^{6}$Leiden University, P.O. Box 9500, 2300 RA Leiden, The Netherlands\\
}

\begin{document}

%\date{Accepted  Received ; in original form }
\date{}

\pagerange{\pageref{firstpage}--\pageref{lastpage}} \pubyear{2013}

\maketitle

\label{firstpage}

\begin{abstract}
We apply a chemical evolution model to investigate the sources and evolution of dust in a sample of 26 high-redshift ($z>1$) submillimetre galaxies (SMGs) from the literature, with complete photometry from ultraviolet to the submillimetre. We show that dust produced only by low--intermediate mass stars falls a factor 240 short of the observed dust masses of SMGs, the well-known `dust-budget crisis'. Adding an extra source of dust from supernovae can account for the dust mass in 19\,per\,cent\, of the SMG sample. Even after accounting for dust produced by supernovae the remaining deficit in the dust mass budget provides support for higher supernova yields, substantial grain growth in the interstellar medium or a top-heavy IMF. Including efficient destruction of dust by supernova shocks increases the tension between our model and observed SMG dust masses. The models which best reproduce the physical properties of SMGs have a rapid build-up of dust from both stellar and interstellar sources and minimal dust destruction. Alternatively, invoking a top-heavy IMF or significant changes in the dust grain properties can solve the dust budget crisis only if dust is produced by both low mass stars and supernovae and is not efficiently destroyed by supernova shocks.
\end{abstract}


\section{Introduction}
\label{Intro}
The first blind submillimetre surveys discovered a population of highly star-forming ($100-1000\,$M$_\odot$yr$^{-1}$), dusty galaxies at high redshift (\citealp*{Smail97}; \citealp{Hughes98, Barger98, Eales99}). These submillimetre galaxies (SMGs) are thought to be undergoing intense, obscured starbursts \citep{Greve05, Alexander05, Tacconi06, Pope08}, which may be driven by gas-rich major mergers \citep[e.g.][]{Tacconi08, Engel10, Wang11, Riechers11}, or streams of cold gas \citep{Dekel09, Dave10, vandeVoort11a}. Observational studies show that SMGs typically have stellar masses of $\sim 10^{11}\,\rm M_{\odot}$ \citep[e.g.][]{Hainline11, Magnelli12}, large dust masses ($\sim 10^{8-9}\,\rm M_{\odot}$; \citealt{Santini10, Magdis12, Simpson14}), high gas fractions ($30-50$\,per\,cent; \citealt{Tacconi08,Bothwell12}) and Solar or sub-Solar metallicities \citep{Swinbank04, Banerji12, Nagao12}.

The source of interstellar dust in SMGs is still a controversial issue, particularly whether it originates from supernovae (SNe) or from the cool, stellar winds of low-intermediate mass stars (LIMS).  Recent work has revealed a \emph{`dust budget crisis'} (\citealp[][hereafter ME03]{ME03}; \citealp*{Dwek07}; \citealp*{MWH10}; \citealp{Michalowski10b, Santini10}; \citealp*{Gall11a}; \citealp{Valiante11}), whereby it is difficult to explain the high dust masses observed in high redshift galaxies through dust from LIMS\footnote{Note that this problem is not limited to high redshift SMGs, and is also seen in galaxies at low redshift \citep[e.g.][]{Matsuura09, Dunne11, Rowlands12, Smith_HRS12, Boyer12}.}. At $z>5$ this is further compounded as there is little time for LIMS to produce significant amounts of dust (ME03; \citealt{Dicris13}). The surprisingly constant dust-to-metals ratio measured in galaxies over a wide range of cosmic time also indicates that a rapid mechanism of dust formation is needed \citep*[][and references therein]{Zafar13}, requiring dust formation timescales to be the same order as the metal enrichment timescale.  Although \citet{Valiante09} and \citet{Dwek11} argue that AGB stars may contribute significantly to the dust budget after only $150-500$~Myrs (and thus may be a significant source of dust at high redshift), the amount of dust produced is highly sensitive to the assumed initial mass function (IMF). Furthermore, in the former study a high star-formation rate (SFR) in excess of $1000\,$M$_\odot$yr$^{-1}$ sustained over 
$\sim 0.3-0.4$\,Gyr is required to build up a significant mass of dust. Due to their short lifetimes, massive-star SNe have long been proposed as a potential source of dust at early times (ME03; \citealp{Nozawa03, Dunne03a, Dwek07, Dunne09b}; \citealp*{Gall11Rev}).

Observational evidence of dust formation in SN ejecta has come to light recently with SN1987A, Cas A and the Crab Nebula remnants containing significant quantities of dust ($0.1-1\,\rm M_{\odot}$; \citealp{Dunne09b,Matsuura11,Gomez12b}).  There is now little doubt that dust is formed in SN ejecta \citep{Dunne03a,Sugerman06, Rho08, Barlow10, Matsuura11, Temim12, Gomez12b} though the amount of dust which will ultimately survive the supernova shocks is still highly uncertain \citep{Bianchi07, Kozasa09, Jones_Nuth11}. Additionally, dust grain growth in the ISM \citep{Draine09} has been proposed as an extra source of dust in galaxies at both high redshift \citep{MWH10,Hirashita_Kuo11,Valiante11,Calura14}, and at low redshift \citep{Dwek07, Dunne11, Inoue12, Kuo_Hirashita12, Mattsson_Andersen12b, Boyer12, Asano13}, which could make up the shortfall in the dust budget of galaxies. The difficulty with determining the origin of dust in galaxies and its lifecycle arises due to a lack of large samples of sources in which to test these issues.  Previous authors including ME03; \citet{Valiante09,Valiante11,Dwek11,Gall11a} and \citet{MWH10} investigated the origin and evolution of dust in high redshift galaxies, but these were limited to one or two (or, at most, a handful) extreme starbursting systems, selected in a non-uniform way and often missing crucial far-infrared (FIR) photometry spanning the peak of the dust emission.

In \citet{Rowlands14b} a sample of SMGs were carefully selected from the comprehensive data in \citet{Magnelli12} and galaxy properties were derived for the population by fitting their spectral energy distributions (SEDs) from the UV to the submillimetre in a consistent way. Here, we investigate the origin of dust in these high redshift SMGs using an updated version of the chemical evolution model of ME03 which incorporates realistic star-formation histories (SFHs) for each galaxy, with greater complexity than previous chemical evolution studies have attempted. The sample properties and derivation of the observational parameters are described in full in \citet{Rowlands14b} (see also \citealt{Magnelli12}, hereafter M12).  We briefly comment on our sample selection and the spectral energy distribution (SED) fitting method in Section~\ref{sec:sample_selection}.  In Section~\ref{sec:chem_ev_description} we present the updated chemical evolution model which follows the build-up of dust over time, with comparison to the observed properties of SMGs in Section \ref{sec:chem_ev}. Our conclusions are summarised in Section \ref{sec:conclusions}. We adopt a cosmology with $\Omega_m=0.27,\,\Omega_{\Lambda}=0.73$ and $H_o=71\, \rm{km\,s^{-1}\,Mpc^{-1}}$.

\section{Deriving physical parameters for SMGs}
\label{sec:sample_selection}

In order to investigate the physical properties of SMGs, we selected a high redshift sample from M12 ranging from $1.0<z<5.3$. Full details of the sample selection and caveats/selection effects are provided in M12 and \citet{Rowlands14b}, but briefly, the SMGs are selected from blank field (sub)millimetre surveys ($850-1200\mu$m) which have robust counterparts identified with deep radio, interferometric submillimetre and/or mid-infrared (MIR) imaging. The SMGs are located in fields which have a wealth of multiwavelength observations (GOODS-N, ECDFS, COSMOS and Lockman Hole), which is required in order to derive statistical constraints on galaxy physical properties using SED fitting.

\subsection{SED fitting}
\citet{Rowlands14b} used a modified version of the physically motivated method of \citet*[][hereafter DCE08\footnote{The \citet*{DCE08} models are publicly available as a user-friendly model package {\sc magphys} at www.iap.fr/magphys/.}]{DCE08} adapted for SMGs to recover the physical properties of the galaxies in our sample.  Briefly, the energy from UV-optical radiation emitted by stellar populations is absorbed by dust, and this is matched to that re-radiated in the FIR.  Spectral libraries of 50000 optical models with stochastic SFHs, and 50000 infrared models, are produced at the redshift of each galaxy in our sample, containing model parameters and synthetic photometry from the UV to the millimetre. The model libraries are constructed from parameters which have prior distributions designed to reproduce the range of properties found in galaxies. The optical libraries are produced using the spectral evolution of stellar populations calculated from the latest version of the population synthesis code of \citet{BC03}. The stellar population models include a revised prescription for thermally-pulsing asymptotic giant branch (TP-AGB) stars from \citet{Marigo_Girardi07}. A \citet{Chabrier03} Galactic-disk Initial Mass Function (IMF) is assumed. The libraries contain model spectra with a wide range of SFHs, metallicities and dust attenuations. 

The infrared libraries contain SEDs comprised of four different dust components, from which the dust mass (\md) is calculated. In stellar birth clouds, these components are polycyclic aromatic hydrocarbons (PAHs), hot dust (stochastically heated small grains with a temperature $130-250$\,K), and warm dust in thermal equilibrium ($30-60$\,K). In the diffuse ISM the relative fractions of these three dust components are fixed, but an additional cold dust component with an adjustable temperature between 15 and 30\,K is added. The dust mass absorption coefficient $\kappa_{\lambda} \propto \lambda^{-\beta}$ has a normalisation of $\kappa_{850}=0.077\,\rm{m}^2\rm{kg}^{-1}$ \citep{SLUGS00a,James02}. A dust emissivity index of $\beta=1.5$ is assumed for warm dust, and $\beta=2.0$ for cold dust. The prior distributions for the dust temperatures are flat, so that all temperatures within the bounds of the prior have equal probability in the model libraries.

\begin{figure*}
\centering
\includegraphics{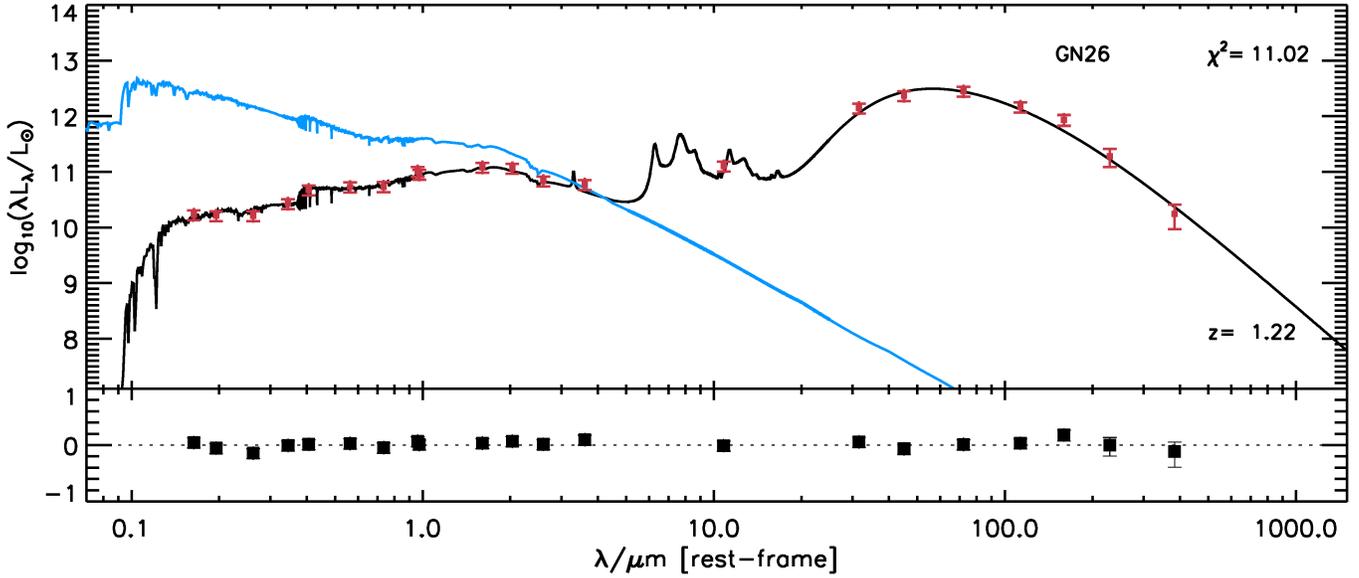}
\caption[Example best-fit rest-frame SED of a high-redshift submillimetre galaxy]{Example best-fit rest-frame SED of a high-redshift submillimetre galaxy, with observed photometry (red points) from the rest-frame UV to the submillimetre. The photometry is described in \citet{Rowlands14b}. The black line is the best fit model SED and the blue line is the unattenuated optical model. The residuals between the best-fit model photometry and the observed data are shown in the bottom panel.}
\label{fig:SED_example}
\end{figure*}

The attenuated stellar emission and dust emission models in the two spectral libraries are combined using a simple energy balance argument, that the energy absorbed by dust in stellar birth clouds and the diffuse ISM is re-emitted in the far-infrared (FIR). Statistical constraints on the various parameters of the model are derived using the Bayesian approach described in DCE08. Each observed galaxy SED is compared to a library of stochastic models which encompasses all plausible parameter combinations. For each galaxy, the marginalised likelihood distribution of any physical parameter is built by evaluating how well each model in the library can account for the observed properties of the galaxy (by computing the $\chi^{2}$ goodness of fit). This method ensures that possible degeneracies between model parameters are included in the final probability density function (PDF) of each parameter. The effects of individual wavebands on the derived parameters are explored in DCE08, and \citet{Smith12}, but we emphasise the importance of using the FIR-submillimetre data from the \emph{Herschel Space Observatory}\footnote{{\it Herschel} is an ESA space observatory with science instruments provided by European-led Principal Investigator consortia and with important participation from NASA.} \citep{Pilbratt10} to sample the peak of the dust emission and the Rayleigh-Jeans slope in order to get reliable constraints on the dust mass and luminosity.

The SEDs of 26 high-redshift SMGs were fitted with {\sc magphys}, producing model parameters for each source including stellar mass ${M}_\ast/$M$_\odot$; dust mass ${M}_\mathrm{d}/$M$_\odot$; dust-to-stellar mass ratio \mdms and the SFR averaged over the last $10^7$ years \sfr/M$_\odot$\,yr$^{-1}$. An example best-fit SED is shown in Fig.~\ref{fig:SED_example}, and the range of values derived for the SMG sample along with their average properties are listed in Table~\ref{tab:smg_observed}. 

\begin{table*}
  \caption{Summary of physical properties for the $z>1$ SMGs derived from stacking the probability density functions (PDFs) derived from {\sc magphys}. For each parameter, we use the first moment of the average PDF to estimate the mean of the population, with the variance on the population taken from the second moment of the average PDF minus the mean squared. The error on the mean is simply the square root of the population variance, normalised by the square root of the number of galaxies in the sample. We also list the full range of median likelihood parameters values. The parameters are: stellar mass $M_\ast/{\rm M_{\odot}}$; dust mass $M_d/{\rm M_{\odot}}$; dust-to-stellar mass \mdms and SFR averaged over the last $10^7$ years.}
\begin{centering}
  \begin{tabular}{ccccc}
\hline
   & ${\rm log}_{10}(\mstar)$ & ${\rm log}_{10}(\md)$ & log$_{10}$(\mdms) & ${\rm log}_{10}({\rm SFR})$  \\ \hline
 Mean &  $10.80 \pm 0.10$  & $9.09 \pm 0.09$ & $-1.71 \pm 0.10$ & $2.59 \pm 0.08$ \\ 
 Range & ($9.87-11.74$) & ($7.89-9.56$) & (-$2.47-$ -$0.81$) & ($1.05-3.34$) \\ \hline 
  \end{tabular}
\end{centering}
\label{tab:smg_observed}
\end{table*}

Evidence from X-ray studies suggest that many SMGs host an active galactic nucleus (AGN) \citep{Alexander05}, indeed six SMGs in the parent sample of \citet{Rowlands14b} show excess emission in the rest-frame NIR, which may be due to dust heated to high temperatures by an obscured AGN \citep{Hainline11}. As the {\sc magphys} SED models do not include a prescription for AGN emission, we follow \citet{Hainline11} by subtracting a power-law component from the optical-NIR photometry of the subset of SMGs which exhibit excess emission in the NIR.  Note that AGN contribute a negligible amount to the flux at wavelengths longwards of rest-frame $30\,\mu$m \citep{Netzer07, Hatziminaoglou10, Pozzi12}. Subtracting the NIR power law results in a reduction in the average stellar mass of the sample by 0.1\,dex and a negligible change in the recent SFR. We have excluded two galaxies in the original M12 sample from our analysis as the uncertainties on the parameters due to the subtraction of the power law are too large.

\begin{figure*}
\includegraphics[width=17.5cm, height=22.8cm]{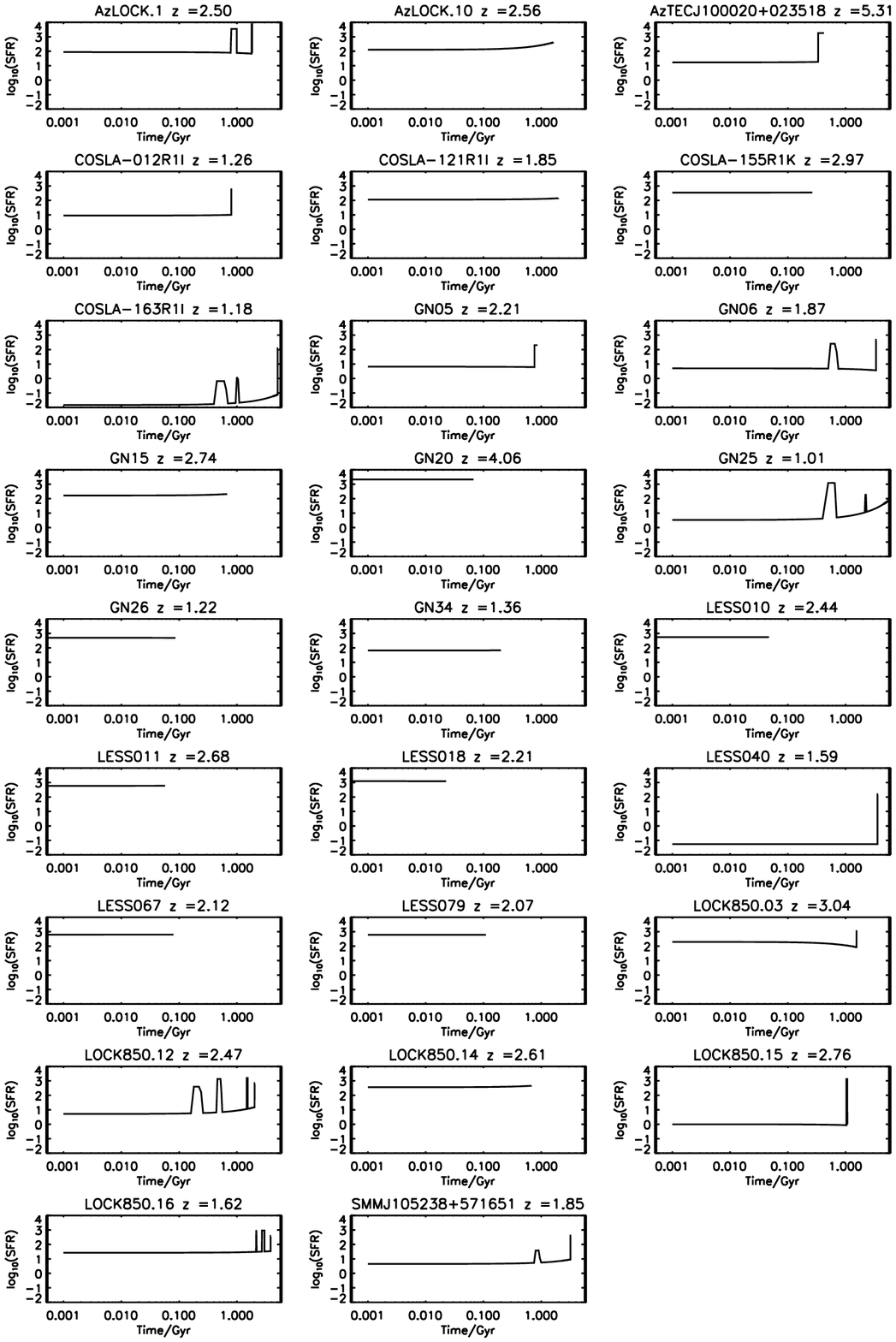}
\caption{Best-fit star-formation histories of the 26 SMGs derived from {\sc magphys} SED fitting (Section~\ref{sec:sample_selection}, see also \citealt{Rowlands14b}).  The majority of star-formation histories can be described as `bursts' of star formation, either because they have a short elevated SFR near the current age, or because their star-formation histories are so short and extreme they can be considered a burst.}
\label{fig:chem_ev_SFHs}
\end{figure*}

{\sc magphys} also allows us to recover an estimate of the star formation history (SFH) for each source. We note that whilst the exact form of the SFH cannot be measured using broad-band SED fitting, the best-fit SFHs are consistent with the physical properties of each SMG. The sensitivity of our results to the SFH is explored in Section~\ref{realisticsfh}. The SFHs in the {\sc magphys} models are parametrised by both exponentially increasing and decreasing models of the form ${\rm exp}(-\gamma t)$, where $\gamma$ is the star-formation time-scale parameter which is distributed uniformly between -1 and 1 Gyr$^{-1}$. The time since the start of star formation in the galaxy ($t$) is uniformly distributed between 0.01\,Gyr and the age of the Universe at the galaxy redshift. Bursts of star formation are superimposed at random times on the underlying SFH, but with a probability such that 50\,per\,cent of the model galaxies will have a burst in the last 2\,Gyr. The strength of the burst is defined as the mass of stars formed in the burst relative to the mass of stars formed in continuous star formation over the lifetime of the galaxy; this parameter ranges from 0.1 to 100. The best-fit SFHs for the SMGs are shown in Fig.~\ref{fig:chem_ev_SFHs}.
Many of the best-fit SFHs are bursty: some galaxies have evidence of recent burst(s) of star formation producing a significant fraction of their stellar mass. Others appear to have a smoother (either exponentially declining or increasing) SFH, though as most of these occur over a short period of time ($10-100$\,Myr), they are also `bursty' in nature. As expected, SMGs are therefore likely to rapidly exhaust their gas supply within $\sim100$ Myr \citep[][and references therein]{Simpson14}.
These SFHs are a key ingredient of our chemical evolution models. It is important to emphasise that unlike most chemical evolution models in the literature, we will not just use simple parametric SFHs, but we will use SFHs that are consistent with the observed galaxy SEDs. 

\subsection{Physical properties of SMGs}
\label{sec:physical}
A detailed comparison of the properties of SMGs and low-redshift dusty galaxies are presented in \citet{Rowlands14b}. In summary, this sample has an average stellar mass of $6.3^{+1.6}_{-1.3}\times 10^{10}\,$M$_\odot$ (in agreement with \citet{Hainline11} and M12) and an average SFR of $390^{+80}_{-70}\,$M$_\odot$yr$^{-1}$ ($\sim$120 times higher than a low redshift galaxy sample matched in stellar mass, where the average SFR $=3.3\pm{0.2}$\,M$_\odot$yr$^{-1}$). This is consistent with the observed evolution in characteristic SFR of galaxies out to $z\sim2$.  The SMGs harbour an order of magnitude more dust with mass $1.2^{+0.3}_{-0.2}\times{10}^9\,$M$_\odot$ compared to $(1.6\pm0.1)\times{10}^8\,$M$_\odot$ for low redshift dusty galaxies selected to have a similar stellar mass. The dust masses derived for the SMGs are consistent with those found in the literature \citep{Santini10, Magdis12, Simpson14}. It is not surprising that a high redshift submillimetre sample would have a higher average dust mass, since moderate dust masses would not be detectable at high redshifts. However, such a selection effect does not account for the much larger space density of galaxies with the highest dust masses at high redshift, since these would have been detected should they exist at lower redshift. This is consistent with the well documented strong evolution in the dust content of massive, dusty galaxies with redshift, in agreement with \citet{DE01, Dunne03b, Eales10_Hermes, Dunne11, Bourne12a} and \citet{Symeonidis13}. 

\begin{table}
\caption{Summary of typical physical properties for SMGs derived from the literature (see main text for details). The parameters are: the final gas fraction $f_{\rm gas}$, metallicity in units of solar metallicity ($Z$; the ratio of metal mass to gas mass, with $Z_\odot=0.019$), the dust-to-metal mass ratio ($\eta_Z = M_{\rm{\rm dust}}/M_Z$) and the gas-to-dust ratio ($\eta_g = M_{\rm{gas}}/M_{\rm{\rm dust}}$).  }
\begin{center}
\begin{tabular}{ccccc}
\hline
   & $f_{\rm{gas}}$ & $Z/$Z$_\odot$ & $\eta_Z$ & $\eta_g$ \\ \hline
 Range & $0.3-0.5$ & $\sim 1$ & $0.5$ & $30-50$ \\ 
\hline 
\end{tabular}
\end{center}
\label{tab:smg_literature}
\end{table}

In order to compare to chemical evolution models we need to know the gas fraction and metallicity of the SMGs. As these values are not available for all SMGs in our sample we compare to literature values derived for similar SMG samples as summarised in Table~\ref{tab:smg_literature}. The gas fractions ($f_g$) of SMGs are, on average, $30-50$\,per\,cent \citep{Tacconi08,Riechers11,Bothwell12} based on CO observations and assuming a conversion from CO luminosity to $M_{\rm{H_2}}$ of $\alpha_{\rm{CO}} = 0.8-1.0 \,\rm{M_{\odot}\,(K\,km\,s^{-1}\,pc^{2})^{-1}}$. 
The typical metallicities ($Z$) of SMGs are found to be Solar or slightly subSolar, albeit with uncertainties due to possible AGN contamination of emission lines \citep{Swinbank04,Banerji12,Nagao12}. The dust-to-metals ratio $\eta_Z$ out to redshift 6 measured from absolute extinction and metal column densities for a sample of $\gamma$-ray burst afterglows and quasar foreground absorption systems is found to be approximately constant (\citealt{Zafar13}).  Typical gas-to-dust ratios ($\eta_g$) for SMGs are estimated at $=46\pm25$ \citep{Swinbank14} assuming an average gas mass of $M_{\rm{H_2}}=(3.6\pm1.0)\times{10}^{10}\,$M$_\odot$ \citep{Bothwell12}. Similar values of $\eta_g=28^{+14}_{-22}$ were found by \citet{Kovacs06}\footnote{All dust masses have been scaled to $\kappa_{850}=0.077\,\rm{m}^2\rm{kg}^{-1}$ used in this work.}.

\section{The Chemical Evolution Model}
\label{sec:chem_ev_description}

In order to investigate the origin of dust in our sample of high redshift galaxies, we compare the observed dust masses of SMGs to predictions using an updated version of the chemical evolution model of ME03. The model is based on chemical evolution models in the literature \citep*{Tinsley1980, Pagel97, Dwek98, Calura08}. By relaxing the instantaneous recycling approximation to account for the lifetimes of stars of different masses, the model tracks the build-up of heavy elements over time produced by stars (LIMS and SNe) where some fraction of the heavy elements will condense into dust. Given an input SFH, gas is converted into stars over time, assuming an initial mass function (IMF). The total mass of the system is given by

\begin{equation}
M_{\rm total}=M_g+M_*,
\label{eq:all}
\end{equation}

\noindent where $M_g$ is the gas mass and $M_*$ is the stellar mass. The gas mass changes with time as described in Eq.~\ref{ginc}, as gas is depleted by the SFR, $\psi(t)$, 
and returned to the ISM as stars die, $e(t)$:
\begin{equation}
{\frac{dM_g}{dt}} = -\psi(t) + e(t) + I(t) - O(t).
\label{ginc}
\end{equation}

\noindent The first two terms in Eq.~\ref{ginc} on their own describe a closed box system, the third term describes gas inflow with rate $I$ and the fourth term describes outflow of gas with rate $O$. Inflows and outflows are discussed further in Sections~\ref{sec:Inflows} \& \ref{sec:Outflows} and are (in the simplest form) parameterized as a fraction of the instantaneous SFR.  Assuming that mass loss occurs suddenly at the end of stellar
evolution, the ejected mass, $e(t)$ from stars is
\begin{equation}
e(t)=\int_{m_{\tau_m}}^{m_U}{\left[m-m_{R}(m)\right]\psi(t-\tau_m)\phi(m) dm},
\label{chemg}
\end{equation}
\noindent and the remnant mass is \[
 m_{R}(m) =
  \begin{cases}
   0.106m+0.446 & \text{if } m\le 8.0\, \rm M_{\odot} \\
   1.5 & \text{if } m> 8.0\, \rm M_{\odot}
  \end{cases},
\] (adapted from \citealt{Prantzos93}). $\tau_m$ is the lifetime of a star of mass $m$ from \citet{Schaller92}, $m_U$ is 100\,M$_\odot$ and $m_{\tau_m}$ is the mass of a star whose age is that of a system where a star formed at $t-\tau_m$ has died at time $\tau_m$.

For consistency with the SED fitting method in \citet{Rowlands14b} we adopt a \citet{Chabrier03} IMF, 
unless stated otherwise. This takes the form:
\[
 \phi_{\rm Chabrier} (m) = 
  \begin{cases}
   0.85 {\rm exp}\left({- \frac{({\rm{log}}(m)-{\rm{log}}(m_c))^2} {2\sigma^2} }\right) & \text{if } m\le 1\, \rm M_{\odot} \\
   0.24 \,m^{-1.3} & \text{if } m> 1\, \rm M_{\odot}
  \end{cases},
\]
\noindent where $m_c=0.079$ and $\sigma=0.69$. The IMF is normalised to $1$ in the mass range $0.1-100$\,M$_\odot$. The choice of a Chabrier IMF results in higher stellar dust production than the Scalo or Salpeter IMFs (i.e. compared to the results in ME03 and \citealt{Dwek98}), since fewer stars with $m<1\,\mathrm{M}_\odot$ are produced in a given population, hence less metals are locked up on timescales of the order of the Hubble time.

The evolution of the mass of metals in the ISM ($M_Z$) is described by 
\begin{equation}
\frac{d(M_Z)}{dt} = -Z(t)\psi(t) + e_z(t) + Z_{I}I(t) - Z_{O}O(t)+ M_{Z,i},
\label{eq:metals}
\end{equation}

\noindent where $Z$ is defined as the fraction of heavy elements {\it by mass} in the gas phase i.e. 
\begin{equation}
 Z = \frac{M_{\rm Z}({\rm gas})}{M_{\rm g}}.
\end{equation}

\noindent The first term of Eq.~\ref{eq:metals} describes the metals locked up in stars, and the second term describes the metals returned to the ISM via stellar mass loss (as described in Eq.~\ref{eq:chemz}). Together these two terms describe the evolution of metals in a closed box system. The third term of Eq.~\ref{eq:metals} describes an inflow of gas with metallicity $Z_{I}$ and the fourth term of Eq.~\ref{eq:metals} describes an outflow of gas with metallicity $Z_{O}$. The final term $M_{Z,i}$ allows for pre-enrichment from Pop III stars.  We set this to zero, but adding pre-enrichment at the expected level of $Z_i \sim 10^{-4}$\,Z$_{\odot}$ \citep[see][for a review]{Bromm_Yoshida11} does not change any of our results.

The mass of heavy elements ejected by stars at the end of their lives is described by 
\begin{eqnarray}
e_z(t)&=& \int_{m_{\tau_m}}^{m_U}\bigl({\left[m-m_{R}(m)\right] Z(t-\tau_m)+mp_z}\bigr)  \nonumber \\ 
 & &\mbox{} \times \psi(t-\tau_m)\phi(m)dm
\label{eq:chemz}
\end{eqnarray}

\noindent where $mp_z$ is the {\it yield} of heavy elements from a star of initial mass $m$ and metallicity $Z$, interpolated from \citet{Maeder92} for massive stars, and \citet{Hoek} for LIMS (for progenitor masses up to 8\,M$_{\odot}$).  The integrated yield ($p_z$) is defined as the mass fraction of stars formed in the mass range $m_1-m_2$ which are expelled in the form of heavy element $z$ in Eq. \ref{eq:Px}, 
\begin{equation}
  p_z=\int_{m_1}^{m_2}{mp_z(m)\phi(m)dm}.
\label{eq:Px}
\end{equation}

The evolution of the dust mass will depend on (i) the IMF (ii) the SFH (iii) the amount of heavy elements produced in stars (the yield) (iv) the dust destruction efficiency and (v) whether or not dust can be formed in the ISM in addition to stellar winds or explosions.  Heavy elements are produced by both LIMS and SNe, therefore in the model, the fraction of metals turned into dust is parameterized by a dust condensation `efficiency' for both SN and LIMS yields. In general, the evolution of dust mass ($M_d$) with time is described by
\begin{eqnarray}
\frac{d(M_d)}{dt}&=&\int_{m_{\tau_m}}^{m_U}\bigl(\left[m-m_{R}(m)\right] 
Z(t-\tau_m)\delta_{\rm lims}+mp_z\delta_{\rm dust} \bigr) \nonumber \\ 
     & &\mbox{} \times \psi(t-\tau_m)\phi(m)dm - (M_d/M_g)\psi(t) \nonumber \\
     & &\mbox{} - M_d\delta_{\rm dest}(t) + M_d\delta_{\rm grow}(t) + M_{d,i} \nonumber \\
     & &\mbox{}+ (M_d/M_g)_{I}I(t) - (M_d/M_g)_{O}O(t).
\label{eq:chemdust}
\end{eqnarray}

The first term within the parentheses describes metals locked up in stars, a fraction of which are then recycled into dust through stellar winds. The second term accounts for dust produced from freshly synthesised heavy elements in stars (LIMS and SNe) with initial stellar mass $1\le m_i\le 40$\,M$_\odot$.  The third and fourth terms account for dust lost in forming stars (astration) and dust lost via destruction processes (parameterized by $\delta_{\rm dest}$, see Section~\ref{Dust_destruction}) respectively.  The fifth and sixth terms represent grain growth in the ISM (parameterized by $\delta_{\rm grow}$, see Section.~\ref{Grain_growth}) and dust produced by Pop III stars (set to zero, see Eq.~\ref{eq:metals}).  The final two terms describe dust mass gained or lost via inflow and outflow of gas i.e. $(M_d/M_g)_I$ and $(M_d/M_g)_O$.

\subsection{Dust Produced by Stars - $\delta_{\rm dust}$}
The dust condensation efficiency in Eq.~\ref{eq:chemdust} ($\delta_{\rm dust}$) describes the fraction of heavy elements which are incorporated into dust for newly synthesised elements.  This can be split into the dust efficiency from supernovae ($\delta_{\rm sn}$) and from the stellar winds of LIMS ($\delta_{\rm lims}$).  

\subsubsection{Dust from LIMS - $\delta_{\rm lims}$}
\begin{figure*}
\includegraphics[trim=0.5mm 0mm 20mm 150mm,clip=true,width=8.7cm]{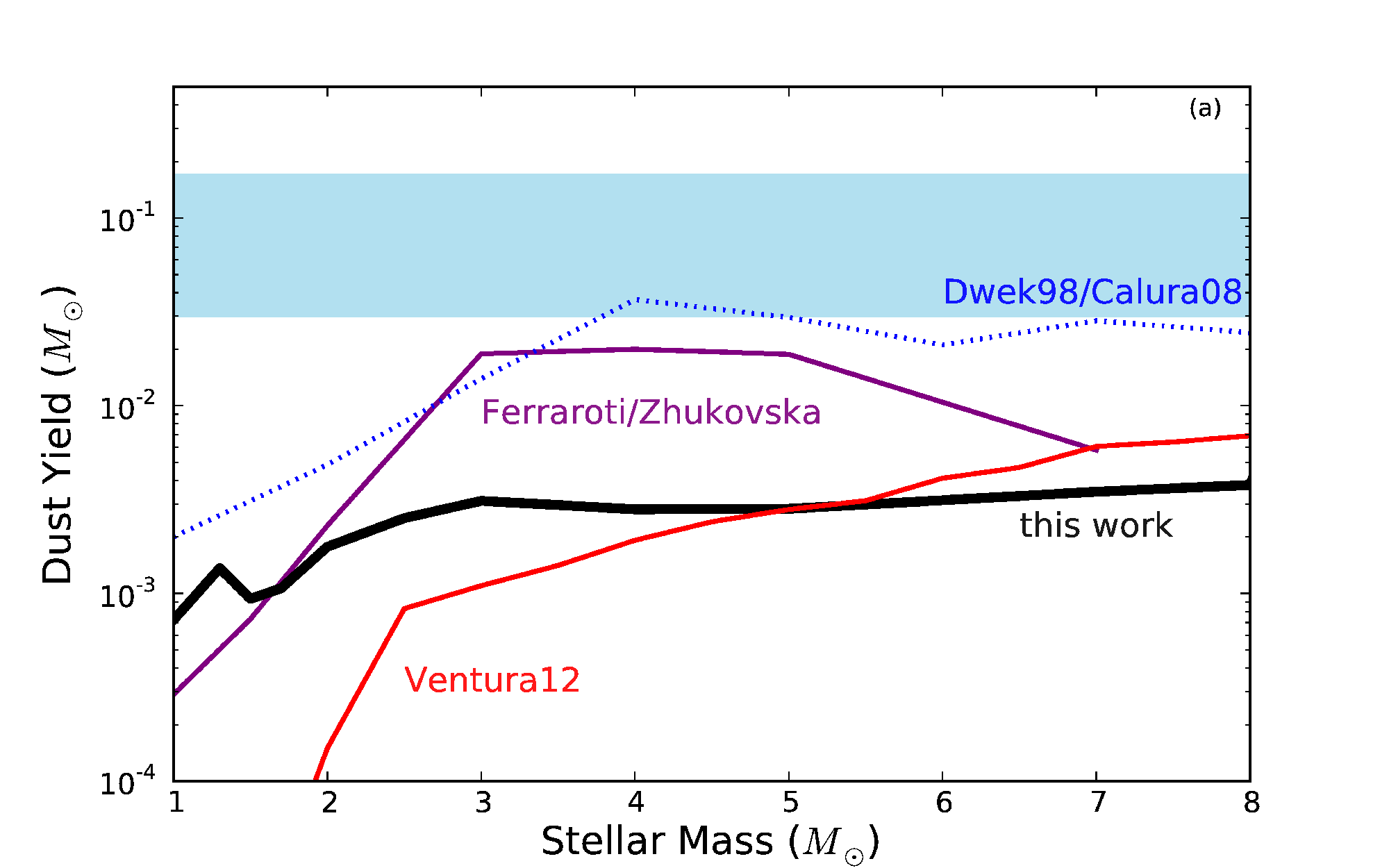}
\includegraphics[trim=0.5mm 0mm 20mm 150mm,clip=true,width=8.7cm]{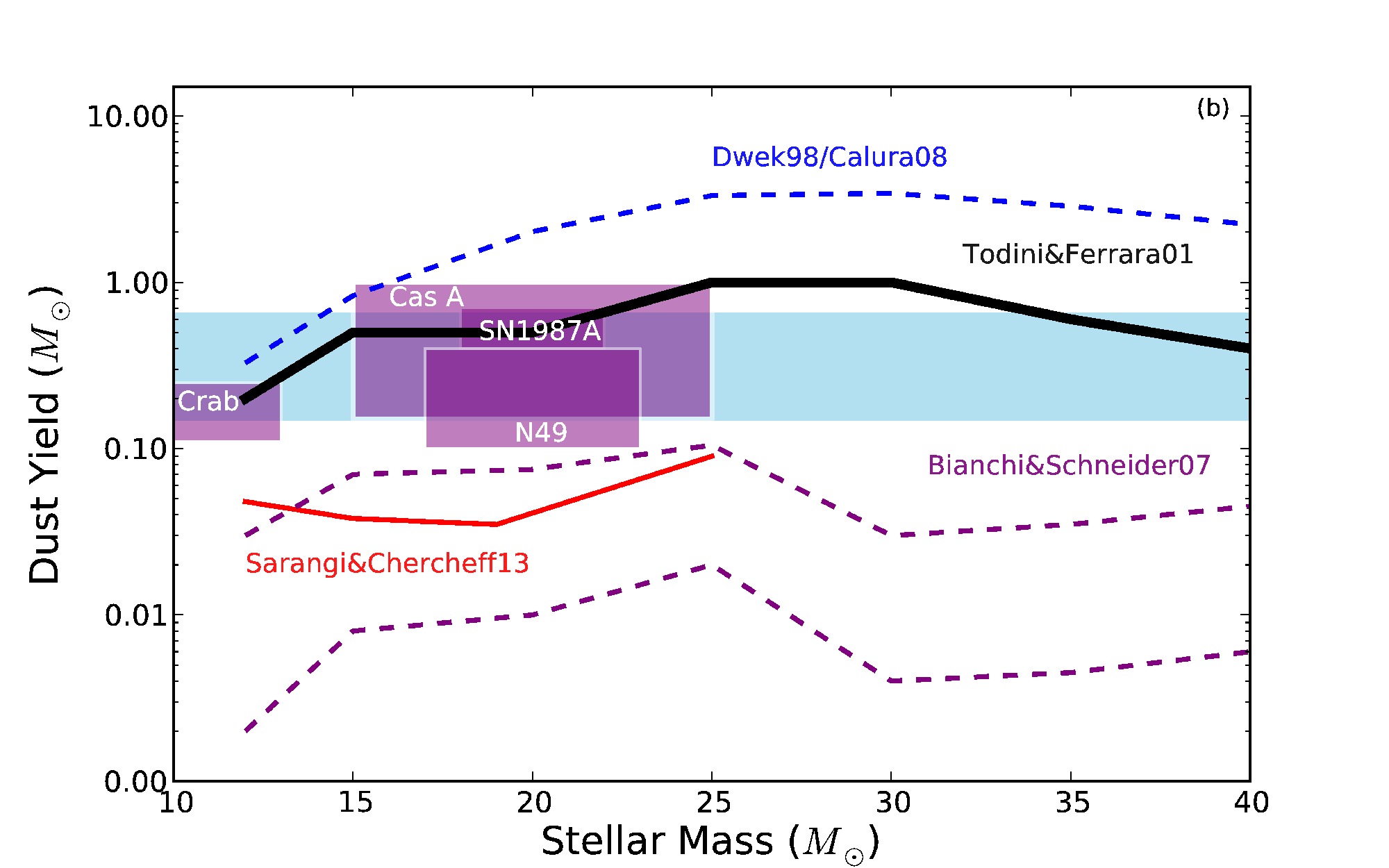}
   \caption{Theoretical and observational dust yields from LIMS and core-collapse SNe at $Z=Z_{\odot}$. {\bf a:} We compare the dust yield from AGB stars from applying the condensation efficiencies in this work and in ME03 ($\delta_{\rm lims}$, black solid line) with the dust yields from \citet{Dwek98} and \citet{Calura08} (blue dotted line).  The purple dashed line shows the yields from the theoretical AGB dust formation models from \citet{Ferrarotti_Gail06} and \citet[][also \citealp{Valiante09}]{Zhukovska08}, with dust masses from \citet{Ventura12} in red. The minimum average dust yield per AGB star required to explain observations of high redshift submillimetre galaxies (\citealp{MWH10}) is shown in the shaded light blue region. {\bf b:}  We compare the theoretical dust yields from core-collapse SNe ($\delta_{\rm SN}$) with observations.  The yields from TF01 used in this work (black solid line), yields from \citet{Dwek98,Calura08} (blue dotted) and \citet{Sarangi13} (red solid); the range of expected yields from the theoretical SN dust formation model of \citet{Bianchi07} which includes dust destruction (purple dashed lines).  The two lines compare the dust mass which has `survived' the SN shock expanding into two different densities.  The average dust yield per SN required to explain high redshift submillimetre galaxies (\citealp{MWH10}, ME03) is shown in the shaded light blue region.  Observed dust masses from Galactic and nearby young SNRs are indicated in the shaded purple regions \citep{Rho08,Dunne09b,Barlow10,Otsuka10,Matsuura11,Gomez12b}.  The boxes indicate the range of dust mass values derived from IR-submillimetre data as well as uncertainties in the mass of the progenitor stars.}
\label{fig:dmstars}
\end{figure*}

In this work, we take the dust condensation efficiencies for LIMS with mass $1 \le m_i\le 8 \,\rm M_{\odot}$ from ME03 and apply them to the metal yields of \citet{Hoek}; the predicted dust yields are then $(1-2000)\times10^{-5}\,\rm M_\odot$ of dust per AGB star (Fig.~\ref{fig:dmstars}a) depending on the initial stellar mass and metallicity. These values are in agreement with subsequent submillimetre observations by \citet{Ladjal10} where dust masses of $(0.01-2000)\times10^{-5}\,\rm M_\odot$ were measured in a sample of AGB stars, and are consistent with the recent AGB theoretical dust formation model of \citet{Ventura12} (Fig.~\ref{fig:dmstars}a, red line). Outside of this mass range, we set $\delta_{\rm lims}=0$.  

Since the amount of dust produced by a population of stars depends not only on the dust condensation efficiency but also the chosen metal yields, and given the wide range of theoretical yields in the literature, here we take a moment to discuss the effect the chosen yields and condensation efficiencies will have on our results.  In Fig.~\ref{fig:dmstars}a, we compare the dust masses from LIMS assumed here (black solid line) to other literature studies.  We compare these with the dust masses from \citet[][also used in \citealp{Calura08}]{Dwek98} where the dust yield is simply assumed to be $1.0 \times m p_Z$ for Mg, Si, S, Ca, Fe and $^{16}$O where $\rm C/O < 1$, and $1.0 \times m p_C$ where $\rm C/O > 1$; with $p_Z$ taken from the metal yields of \citet{Renzini81}.  This rather unrealistic assumption therefore assumes that {\it 100 per cent of the available carbon or oxygen formed in LIMS will condense into dust}.   It is clear that for all progenitor masses in the range $1<m_i<5\,\rm M_{\odot}$ the dust masses from LIMS in \citet{Dwek98} are an order of magnitude higher than the dust masses predicted by this work and by \citet{Ventura12}.  The difference stems from the high dust condensation efficiency assumed in \citet{Dwek98} and (ii) the metal yields used.   The \citet{Hoek} theoretical yields used in this work are more physical than those of \citet{Renzini81}, and importantly include the thermally-pulsating AGB phase consistent with our SED fitting technique (see the comprehensive description in \citet{Romano10} for a detailed comparison of the available yields in the literature).  

The oft-used theoretical AGB dust formation model of \citet{Ferrarotti_Gail06} (see \citealt{Zhukovska08}) also predicts higher dust yields from LIMS than this work, particularly in the range $2<m_i<6\,\rm M_{\odot}$; indeed using the Ferraroti \& Gail dust yields (Fig.~\ref{fig:dmstars}a) in our model would lead to 3.6\,times more dust from LIMS within 0.1\,Gyr.  This disagreement is a combination of a choice of (different) input yields and a higher equivalent `$\delta_{\rm lims}$'. We note that the Ferraroti \& Gail dust yields (derived from synthetic stellar evolution models) also disagree with recent results from the self-consistent full stellar evolution model in \citet{Ventura12}.  In some cases, the Ferraroti \& Gail dust yields for a given AGB star exceeds the total yield of heavy elements produced by the van den Hoek \& Groenewegen yields at solar metallicity.
 
Finally, we have compared our choice of yields \citep{Hoek} with \citet{Marigo00} and updated LIMS models from \citet{Karakas10} (which as discussed in \citealt{Romano10} provides the best fit to a range of observations) and find that the \citet{Hoek} yields are often lower in the $3<m_i<6$ progenitor mass range.  However, we argue in favour of keeping these yields since they are the only uniformly calculated set that includes the super-AGB phase between $5-8\,\rm M_{\odot}$ (important for producing much of the chemical enrichment in galaxies, \citealt{Romano10}), they still explain many of the observational tests which are well-fit by the yields from Karakas \citep{Romano10} and they agree with the updated AGB dust model of \citet{Ventura12}. Ultimately, the lower dust yields from LIMS used here compared to \citet{Dwek98,Calura08} and \citet{Ferrarotti_Gail06} are compensated for by our choice of Chabrier IMF compared to the Scalo/Salpeter IMFs used in these studies since the Chabrier IMF produces approximately 4 times more interstellar metals (and therefore dust) after 0.5\,Gyr of galactic evolution compared to Scalo/Salpeter. Note that if we were to use the Ferraroti \& Gail models combined with the Chabrier IMF, we would still not be able to solve the dust budget crisis. Thus the major results of this work are robust to changing the stellar yields from LIMS i.e. our conclusions would not be significantly different if we used the same yields as previous literature studies, though Fig.~\ref{fig:dmstars}a suggest previous works may have overestimated the contribution from LIMS.

\subsubsection{Dust from supernovae - $\delta_{\rm sn}$}
The dust masses formed per core-collapse SN ($m_d = mp_z \delta_{\rm sn}$) are taken from the theoretical model by \citet[][hereafter TF01]{TF01} who predict $\sim 0.1-1.0$\,M$_\odot$ of dust per SN, depending on the progenitor mass and metallicity. Table~\ref{sn_yields} (see also Fig.~\ref{fig:dmstars}b) lists these dust masses for each massive-star SN at solar metallicity for the progenitor range $11<m_i<25\,\rm M_{\odot}$. Table~\ref{sn_yields} also compares the TF01 dust masses with the metal yields per SN expected from stellar evolutionary models in the literature \citep[e.g.][]{Woosley95,Nomoto06} with indicative values for $\delta_{\rm sn}$ for each stellar mass, to allow for comparison with the model.  Note the wide range of predicted stellar yields from SNe and the range of expected dust masses, this makes $\delta_{\rm sn}$ difficult to pin down observationally.

\begin{table*}
\begin{tabular}{cccccc}\hline 
\multicolumn{6}{c}{Theoretical supernova dust and metal yields from core-collapse supernovae} \\ \hline
$m_i$ & $m_d$ (TF01) & $mp_z^a$ (WW95) & $mp_z$ (N06) & $mp_z$ (M92) & $\delta_{\rm sn}$ \\ 
($\rm M_{\odot}$) & ($\rm M_{\odot}$)& ($\rm M_{\odot}$) & ($\rm M_{\odot}$)  & ($\rm M_{\odot}$)& \\ \hline
11 & .. & 0.4 & .. & .. & .. \\ 
13 & 0.2 & 0.8 & 0.7 & .. & 0.2--0.3 \\ 
15 & 0.5 & 1.4 & 0.6 & 1.32 & 0.4--0.8 \\ 
20 & 0.5 & 3.0 & 2.1 & 2.73 & $\sim$ 0.2 \\ 
25 & 1.0 & 5.1 & 4.3 & 4.48 & $\sim$ 0.2 \\ \hline
\multicolumn{6}{c}{Observed supernova dust yields and estimated condensation efficiencies} \\ \hline
$m_i$ & $m_d$ & $\delta_{\rm sn}$ (WW95) &$\delta_{\rm sn}$ (N06)  &$\delta_{\rm sn}$ (M92) &\multicolumn{1}{l}{Source} \\  
($\rm M_{\odot}$) & ($\rm M_{\odot}$) &  &&  & \\ \hline 
8--13 & 0.1--0.2 & 0.3--0.5 & .. & .. & \multicolumn{1}{l}{Crab Nebula} \\
15--25 & 0.1--1.0 & 0.1--0.7& 0.2--1.0 & 0.02-0.76 &\multicolumn{1}{l}{Cassiopeia A} \\
.. & 0.1--0.4 & 0.02--0.3 & 0.02--0.7 & 0.02-0.76& \multicolumn{1}{l}{N49$^b$} \\
20 & 0.4--0.7 & 0.1--0.2 & 0.2--0.3& 0.15--0.26& \multicolumn{1}{l}{SN~1987A$^b$} \\  \hline
\end{tabular}
\caption{A list of the predicted dust mass $m_d$ from \citet{TF01} produced in each supernova event for different stellar progenitor mass $m_i$ at $Z_{\odot}$.  Also shown are the {\it total} metal yields ($mp_z$) available in the ejecta (WW95:~\citealt{Woosley95}; N06:~\citealt{Nomoto06}; M92:~\citealt{Maeder92}) and an estimate of the SN dust efficiency parameter in the model $\delta_{\rm sn} = m_d/(mp_z)$ (the range indicates the difference in yields WW95, N06, M92)$^a$.  Observational values for $m_d$ from the literature are also shown; note that only SNRs with {\it Herschel} observations have been included here since SN dust masses estimated from mid-infrared photometry of remnants are likely to be lower limits.
$^a$ The metal mass will be overestimated since this includes all metals produced except for light elements and it is likely that the metals available to form dust is less (depending on the carbon/oxygen ratio and compounds formed).
$^b$ The Large Magellanic Cloud remnants (N49, SN~1987A) are in a lower metallicity environment (0.5\,Z$_{\odot}$) compared to the Galactic SNRs, but the predicted metal and dust yields for these SNe do not change significantly between $0.1-1.0$\,Z$_{\odot}$.}
\label{sn_yields}
\end{table*}

It has only recently become possible to compare total theoretical dust masses from SNe with observations. FIR and submillimetre observations with {\it Herschel} have detected cool ($T_{d} \sim 30-40$\,K) dust in SNRs with masses of $\sim 0.1\,\rm{M}_{\odot}$ \citep{Rho09, Barlow10, Gomez12b}.  There is also evidence that the Cas A and SN1987A SNRs have a more massive colder population of dust ($T_{d}\sim20$\,K, $m_d \sim 0.4-1.0\rm \,M_{\odot}$, \citealt{Dunne03a,Dunne09b, Matsuura11}), and these dust masses are close to the higher end of the range predicted by TF01 (see Table~\ref{sn_yields} and Fig.~\ref{fig:dmstars}).  However, little is known about how much dust will survive the passage through the shockfront \citep*[e.g.][]{Bianchi07}.  Assuming dust destruction in the reverse shock is not efficient, the TF01 model dust yields appear to explain the highest observed dust masses in nearby SNRs at solar metallicity (Table~\ref{sn_yields}, Fig.~\ref{fig:dmstars}b).   Therefore we set $\delta_{\rm sn}$ to reproduce the TF01 models in the mass range $9\le m_i \le 40 \,\rm M_{\odot}$ (elsewhere it is set to zero).  In Fig.~\ref{fig:dmstars}(b) we compare the SN dust masses predicted per progenitor star used here (based on TF01 and {\it Herschel} observations) with those used in other chemical evolution studies, in particular the works of \citet{Dwek98} and \citet{Calura08} who assume dust condensation efficiencies of $0.8\times mp_Z$ for Mg, Si, S, Ca, Fe and $^{16}$O and $0.5 \times mp_C$ (where $p_C$ and $p_Z$ is taken from the published yields by \citealp{Woosley95}). We also compare the range of expected SN dust yields from the dust formation model in \citet{Bianchi07}, an updated version of TF01, which also includes destruction by supernova shock waves expanding into the surrounding ISM.  The range of dust masses obtained from FIR/submillimetre observations of Galactic and nearby SNRs are indicated via the shaded purple boxes.

The average dust yield per SNe required to explain dust masses in a handful of high-redshift SMGs (from \citet{MWH10} and ME03) is also highlighted by the light blue shaded region on this plot. The highest observed dust masses from {\it Herschel} and the TF01 models agree well with the their estimates. Finally, we ignore dust formation in Type Ia supernovae as recent {\it Herschel} observations suggest these events are not contributing a significant mass of dust to the ISM \citep{Gomez12a}.

\subsection{Dust Destruction - $\delta_{\rm dest}$}
\label{Dust_destruction_intro}

Since dust is thought to be removed from the ISM by the sputtering and shattering of dust grains by supernova shocks \citep{McKee89}, it needs to be accounted for in this model. The efficiency of dust destruction is highly uncertain but is assumed to depend on the density and composition of the ISM, and the supernova shock velocity \citep{McKee89, Jones04, Dwek07}. Depending on the adopted destruction efficiency, the predicted dust mass in a galaxy from chemical evolution models can vary by a factor of 10 for a Salpeter IMF \citep{Dwek07}. We follow \cite{Dwek07} by parameterising the dust destruction as a function proportional to the SN rate. In this case, the dust destruction timescale $\tau_{\rm dest}$ is described by Eq.~\ref{eq:Dust_destruction}:
\begin{equation}
\tau_{\rm dest} = {M_g  \over{m_{\rm ISM} R_{\rm SN} (t)}}
\label{eq:Dust_destruction}
\end{equation}

\noindent where $M_g$ is the gas mass and $R_{\rm SN}$ is the SN rate:
\begin{equation}
R_{\rm SN} (t) = \int^{40\,\rm M_{\odot}}_{8\,\rm M_{\odot}} \phi(m) \psi\left(\tau - \tau_m \right)dm.
\label{eq:snrate}
\end{equation}

The parameter $m_{\rm ISM}$ is the effective mass of ISM cleared by each SN event, usually assumed to be $1000\, \rm M_{\odot}$ for typical Galactic interstellar densities of $0.1-1\,\rm cm^{-3}$ \citep[e.g.][see Section~\ref{Dust_destruction} for more details]{Dwek11,Gall11a}. The dust destruction timescale varies over time depending on the gas fraction and supernova rate, with a value of $<0.09\,$Gyr for a gas fraction of 0.5 and SFR of $>60\,$M$_\odot$yr$^{-1}$.
The destruction parameter in our model is given by $\delta_{\rm dest} = \tau_{\rm dest}^{-1}$. 

\subsection{Dust Growth - $\delta_{\rm grow}$}
\label{Grain_growth_intro}
We also include a prescription which accounts for accretion of atoms onto dust grain cores in the cold, dense regions of the ISM \citep{DwekScalo80, Tielens98, Draine09}.  The shortfall of dust from stellar sources in galaxies supports evidence that  grain growth is a significant contributor to the dust budget \citep[e.g.][]{Draine09, Zhukovska08, Michalowski10b, Pipino10, Gall11Rev, Dunne11,Jones_Nuth11,Valiante11, Kuo_Hirashita12,Mattsson_Andersen12b,Boyer12,Asano13}. We follow the prescription of \citet{Mattsson_Andersen12b}, where the rate of grain growth is linked to the metallicity (as grains accrete metals in order to grow), and the SFR, assumed to be proportional to the amount of molecular gas in a galaxy since dense regions (where molecular gas is present) is also the environment where grain growth is likely to take place.
The timescale for grain growth $\tau_{\rm grow}$ in the ISM is then given by \citet{Mattsson_Andersen12b}

\begin{equation}
\tau_{\rm grow} = \tau_{o} \left(1 - \frac{\eta_d}{Z}\right) ^{-1},
\label{eq:graingrowth}
\end{equation}

\noindent where $\eta_d$ is the dust-to-gas ratio and $Z$ is the metallicity. $\tau_0$ is defined in \citet{Mattsson_Andersen12b} as:

\begin{equation}
\tau_{o}^{-1} = \frac{\epsilon Z}{M_g} \times~\psi,
\end{equation}

\noindent where $\psi$ is the SFR and $\epsilon$ is an efficiency parameter (which is unconstrained). The growth parameter in our model is given by $\delta_{\rm grow} = \tau_{\rm grow}^{-1}$ (see Section~\ref{Grain_growth} for more details). We define $\tau_{o}^{-1}$ as proportional to the SFR because it is mathematically convenient, but we note that in a more correct physical model of $\tau_{o}^{-1}$ may be proportional to the molecular gas density. Note that the parameterisation of the grain growth does not make much difference in the present context.

\subsection{A more realistic treatment of star-formation history}
\label{realisticsfh}
The detailed treatment of the lifetimes of stars of different stellar masses is important for SMGs which may have bursts of star formation which occur on short timescales.  Previous studies of chemical evolution in SMGs (e.g. ME03) often assumed a SFR proportional to the gas mass which decreased smoothly with time. One of the main differences between this work and ME03 (among others) is the incorporation of a more realistic SFH with bursts of variable strength and duration, and with an underlying SFH which can be either exponentially rising or declining (indeed 15 of our SMGs have exponentially rising SFHs, see Fig.~\ref{fig:chem_ev_SFHs}). Incorporating the SED-derived SFHs allows us to carry out chemical evolution modelling in a manner that is consistent with the SED fitting method.

Degeneracies between the {\sc magphys}-derived SFH parameters, such as the timing, strength and duration of bursts, means that the best-fit SFH may not be a unique solution. Since the SFHs are constrained predominantly by the UV-optical light which is emitted mostly by stars $<100$\,Myrs old, there is a large uncertainty on the form of the SFH at times prior to this. Our approach, however, is correct in the statistical sense for the SMG sample (if not for each individual galaxy), such that each best-fit SFH produces the observed UV-submillimetre SEDs of these galaxies. Whilst differences between the SFHs can cause some variation in the dust mass, the overall mass of dust produced will ultimately depend on the mass of metals formed, which is governed by the total mass of stars formed \citep{Edmunds_Eales98}. The adopted SFH gives a physically plausible and self-consistent representation of the SFH which we can use as an input to our chemical evolution models. Although a small number of studies have used physically realistic SFHs consistent with galaxy physical properties \citep{Valiante09, Valiante11}, our method represents an improvement over previous works that use arbitrary SFHs (or indeed just one SFH to describe all galaxies) which may not be appropriate. 

In addition to the derived SFHs for each individual SMG, we also use four fiducial models to explore the changes in dust mass when key parameters such as the SFH and IMF are varied. These fiducial models represent the whole range of continuous SFHs as derived by the {\sc magphys} SED-fitting and thus provide a test of the range of entire dust masses expected to be formed. The fiducial models are parametrised by (i) an exponentially declining SFH with initial SFR of $150$\,M$_\odot$yr$^{-1}$ (ii) an exponentially increasing SFH with final SFR of $150$\,M$_\odot$yr$^{-1}$, (iii) a constant SFR of $150$\,M$_\odot$yr$^{-1}$, and (iv) a burst SFH with all the stellar mass produced in a burst of $1000$\,M$_\odot$yr$^{-1}$. Each model has an initial gas mass set to the median initial gas mass of the SMG sample (see Section~\ref{sec:chem_ev}). The fiducial SFHs reach the mean stellar mass of the SMG sample (Section~\ref{sec:chem_ev}) at 0.9, 4.4, 0.6 and 0.1 Gyr after the onset of star formation, respectively.  The short-burst fiducial SFH produces a much lower mass of dust at a given gas fraction compared to the other three SFHs.  In total, the variation between the dust masses built up by the model for the different fiducial SFHs is a factor of 29 (if dust is formed only by LIMS) and a factor 1.5 if dust is contributed by both SNe and LIMS.

\section{The origin of dust in SMGs}
\label{sec:chem_ev}

We now explore how well different chemical evolution models can reproduce the dust masses of our SMG sample, and other observational properties of SMGs as a population (Section~\ref{sec:physical}, Table~\ref{tab:smg_literature}). To model these sources, in the first instance, we consider a closed box model, assuming no inflow or outflow of gas or metals. The initial gas mass is set at $2\times$ the best-fit stellar mass derived from the SED fitting, such that at the end of the SFH history, $\sim50$\,per\,cent\, of the total galaxy mass ends up in stars\footnote{We use the best-fit {\sc magphys} stellar mass to be consistent with the best-fit SFH.}. The initial gas masses for this sample (with median $M_g(0) = 1.25 \times 10^{11}\,\rm M_{\odot}$) are therefore tuned to reproduce the observed gas fractions of SMGs.  In the closed box model, the average final gas mass of the SMGs ($5.8\times10^{10}\,$M$_\odot$) is in agreement with observations. Furthermore, by design, the final stellar masses are in close agreement with the best-fit stellar masses (mean of the PDF is $6.3^{+1.6}_{-1.3}\times 10^{10}\,\rm M_{\odot}$) derived from the SED fitting (Table~\ref{tab:smg_observed}; \citealp{Rowlands14b}). 

We then use our model to go beyond the simple closed box by including the effects of inflows and outflows on the gas, metals and dust. We also explore the effect of dust destruction and grain growth on the dust mass. A summary of all of the observational results derived from the different chemical evolution models considered in this work are given in Table~\ref{tab:chem_ev_summary_all}, with the `final' dust masses summarised in Table~\ref{tab:mdust_summary}.  Table~\ref{tab:chem_ev_summary} shows a set of `good' models which reproduce many of the observed properties of the SMG sample. The build-up of dust and stellar mass over time for different chemical evolution models is shown for each individual SMG in Fig.~\ref{fig:chem_ev_Mstar_Mdust}.

\begin{figure*}
\includegraphics[width=1.0\textwidth, height=20cm]{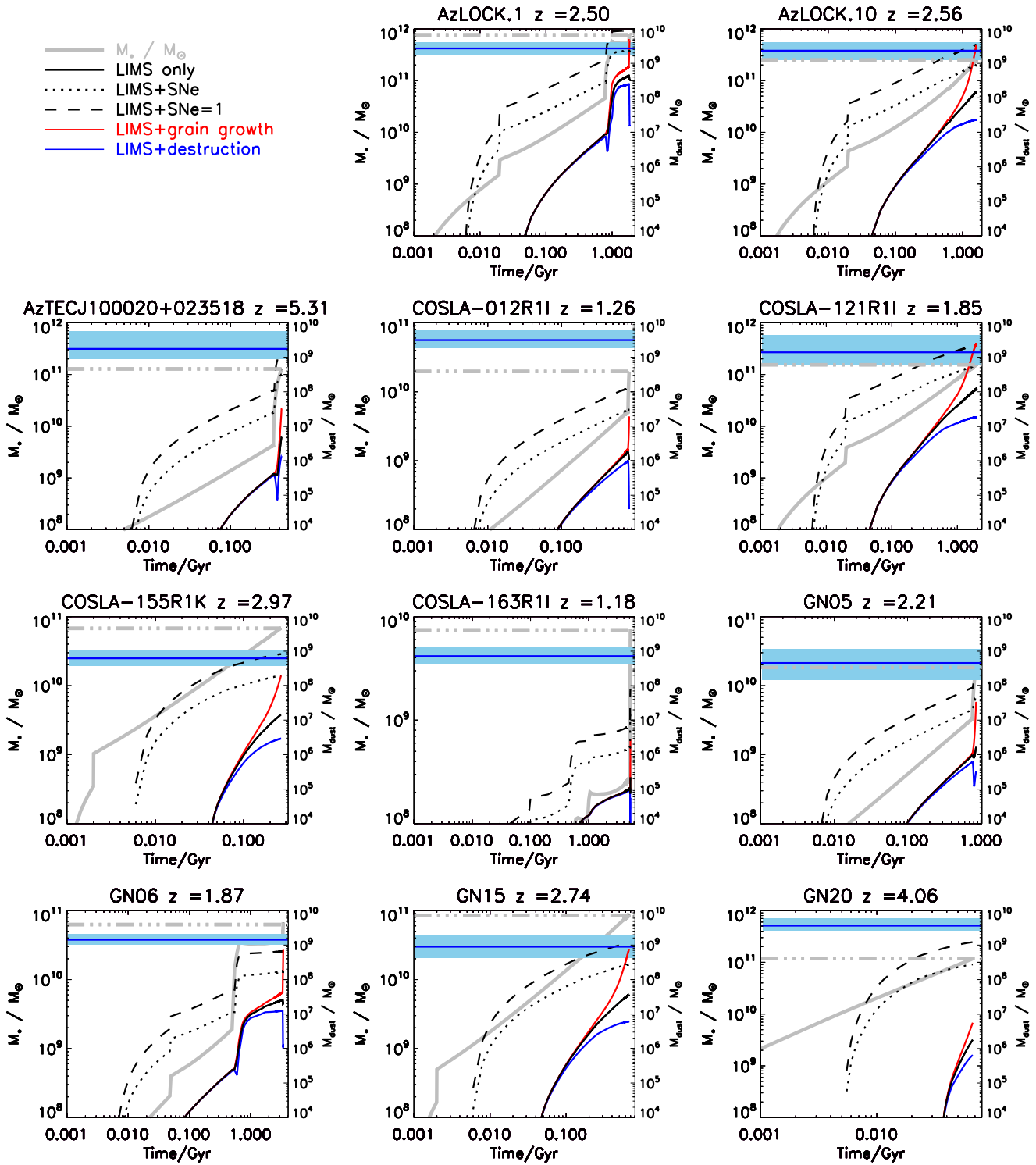}
\caption[The stellar and dust mass evolution over time for submillimetre galaxies derived from chemical evolution modelling.]{The stellar and dust mass evolution over time derived from chemical evolution modelling for the sample of 26 SMGs. Note that the y-axis labels of each panel are different on the right and left sides. The stellar mass growth from the input {\sc magphys} SFH is represented by the grey line and corresponds to the left axis. All of the other lines represent different dust models and correspond to the right axis. The black solid line is the dust mass produced by low--intermediate mass stars (LIMS) only, the black dotted line is LIMS and supernova dust, and the black dashed line is LIMS and maximal supernova dust production. The red line represents the dust mass in a model where dust is produced by LIMS and grain growth, and the blue line shows the dust mass if dust produced by LIMS is destroyed by supernova shocks. At early times, dust destruction and grain growth models have a dust mass track similar to that with dust from LIMS only. Horizontal dot-dashed grey and blue lines represent the observed best-fit stellar masses and median-likelihood dust masses, respectively, with the blue shaded region indicating the 84th--16th percentile range from the SED fitting.}
\label{fig:chem_ev_Mstar_Mdust}
\end{figure*}

\begin{figure*}
\includegraphics[width=1.0\textwidth, height=24cm]{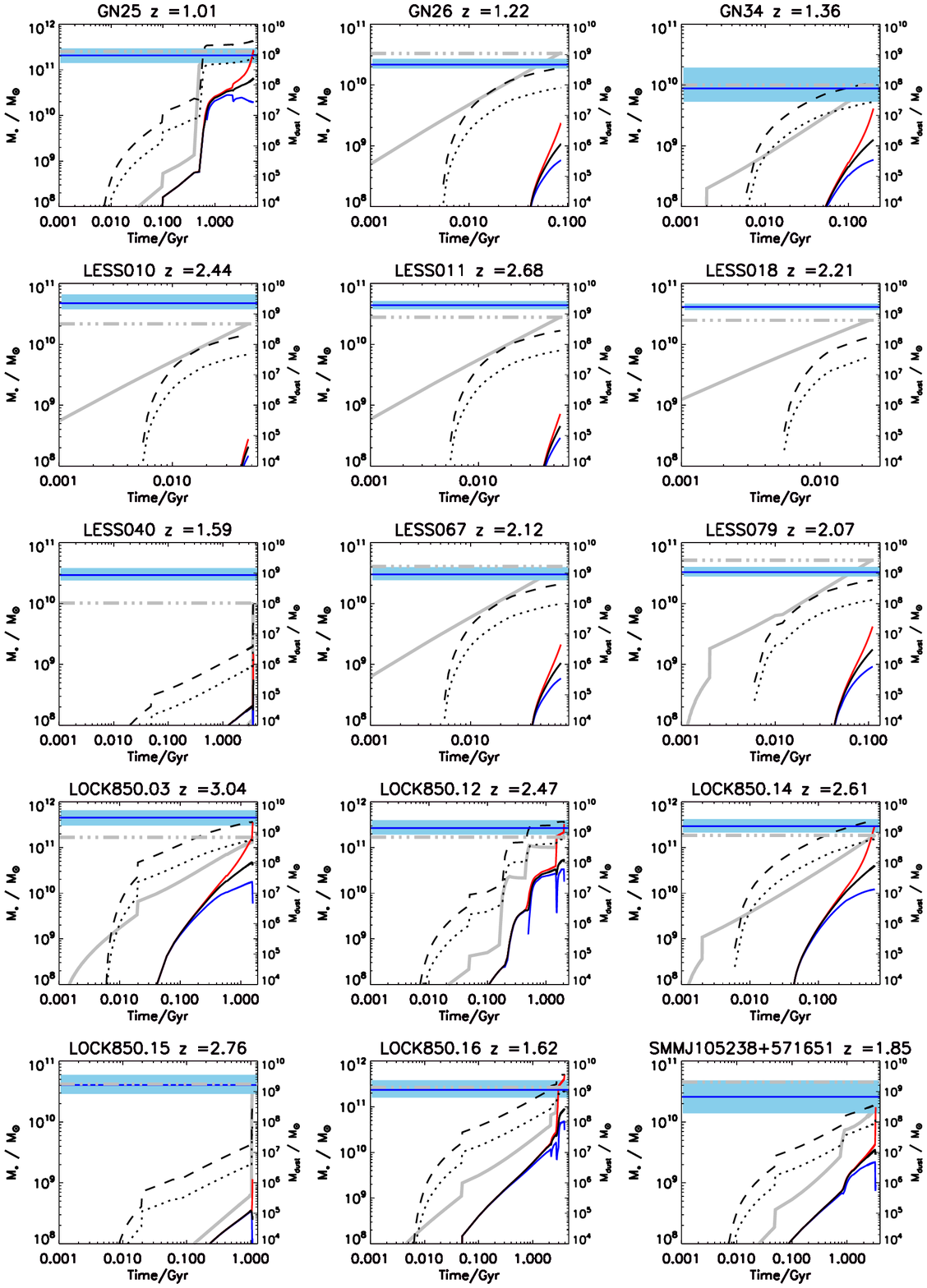}
\contcaption{}
\end{figure*}

\begin{table}
\begin{center}
\begin{tabular}{ >{\raggedright\arraybackslash}m{3.5cm} >{\centering\arraybackslash}m{2.0cm} >{\centering\arraybackslash}m{2.0cm}}
\hline
Model & log$_{10}(M_{\rm{\rm d}}$/M$_\odot$) & Sample percentage \\
\hline
\vspace{0.1cm}
LIMS dust only & 6.70 &   0 \\ \vspace{0.2cm}
LIMS+supernova dust & 8.27 &  19 \\ \vspace{0.2cm}
LIMS+maximal supernova dust & 8.88 &  58 \\ \vspace{0.2cm}
LIMS only+destruction ($\mISM=1000\,$M$_\odot$) & 5.94 &   0 \\ \vspace{0.2cm}
LIMS only+destruction ($\mISM=100\,$M$_\odot$) & 6.61 &   0 \\ \vspace{0.2cm}
LIMS+supernova dust+destruction ($\mISM=1000\,$M$_\odot$) & 7.25 &   0 \\ \vspace{0.2cm}
LIMS+supernova dust+destruction ($\mISM=100\,$M$_\odot$) & 8.07 &  12 \\ \vspace{0.2cm}
LIMS+grain growth & 7.53 &  42 \\ \vspace{0.2cm}
LIMS+supernova dust+grain growth & 8.99 &  62 \\ \vspace{0.2cm}
LIMS + maximal supernova dust + grain growth & 9.06 &  69 \\ \vspace{0.2cm}
LIMS+SNe+destruction ($\mISM=1000\,$M$_\odot$) + grain growth & 7.78 &   0 \\ \vspace{0.2cm}
LIMS+destruction ($\mISM=1000\,$M$_\odot$) +grain growth & 6.37 &   0 \\ \vspace{0.2cm}
LIMS inflow ($I=1\times$\,SFR) & 6.80 &   0 \\ \vspace{0.2cm}
LIMS outflow ($O=1\times$\,SFR) & 6.54 &   0 \\ \vspace{0.2cm}
LIMS inflow+outflow ($I=O=1\times$\,SFR) & 6.59 &   0 \\ \vspace{0.2cm}
LIMS dust only ($2\times$ initial gas mass) & 6.59 &   0 \\ \vspace{0.2cm}
LIMS+SNe+destruction ($\mISM=100\,$M$_\odot$) +grain growth & 8.90 &  54 \\ \vspace{0.2cm}
LIMS+SNe+destruction ($\mISM=100\,$M$_\odot$) +grain growth + inflow ($I=1\times$\,SFR) outflow ($O=1\times$\,SFR) & 8.52 &  35 \\
\hline
\end{tabular}     
\caption{Summary of the median dust masses predicted by different chemical evolution models for the 26 SMGs. For reference the average observed dust mass in the SMG sample is $10^{9.09}$\,M$_\odot$.  Column 3 lists the percentage of the sample of SMGs for which the model reproduces the observed dust masses.}
\label{tab:mdust_summary}
\end{center}
\end{table}

\subsection{Dust production in Stars - LIMS only}
In the first instance we consider dust production in a closed box from LIMS only for each SMG in the sample, assuming no dust destruction as an optimistic case.  The dust produced by LIMS only is indicated by the solid black line in Fig.~\ref{fig:chem_ev_Mstar_Mdust}. The delay between the onset of star formation (as traced by the build-up of stellar mass shown as the solid grey line) and significant dust production by LIMS is evident in these plots, requiring more than a few hundred Myr to reach dust masses greater than $10^8\,$M$_\odot$. In Fig.~\ref{fig:SFR_Mdust_Magnelli}~(a) we show the difference between the median likelihood dust mass from the SED fitting and the final dust mass derived from the chemical evolution modelling. The dust masses calculated from the chemical evolution model with dust from LIMS fall far short of the observed dust masses for the majority of SMGs. On average, the theoretical dust masses are $5.0\times{10}^6$\,M$_\odot$, which is a factor 240 lower than the average observed dust mass in the SMG sample ($1.2^{+0.3}_{-0.2}\times{10}^9\,$M$_\odot$). This provides definitive evidence that (without changing the IMF) the majority of dust in SMGs must come from a source other than LIMS.  Although noted by previous authors, this was based previously on smaller samples and/or simple parameterized SFHs (e.g. ME03, \citealp{Matsuura09,MWH10,Gall11Rev,Dunne11}). Whilst different SFHs can produce differences in the dust mass of up to a factor of 29 (Section~\ref{realisticsfh}), the dust deficit seen here is far larger than this, and occurs for all SMGs in our sample regardless of the SFH. Thus, this conclusion is robust for our ensemble of SMGs.

For the closed box model, the median metallicity for the SMGs reaches $0.9$\,Z$_\odot$; this will be discussed further in Section~\ref{sec:Inflows}.  The median fraction of metals in the ISM in the form of dust ($\eta_Z = M_d/M_Z$) in the LIMS-only model is 0.4\,per cent; this is well below the dust-to-metal ratios observed in local galaxies and out to redshifts of 6 \citep{Whittet92,Pei99,James02,Watson11,Zafar13} which are typically 50\,per cent.

The closed box model reproduces the observed average gas fraction in SMGs (30--50\,per\,cent), but in reality in a sample of galaxies there will be a range of gas fractions. We therefore briefly explore the effect of increasing the initial gas mass in each galaxy by a factor of two to $M_g(0) = 2.5\times{10}^{11}$\,M$_\odot$. This results in a decrease in the median final dust mass by a factor of 1.3 to $\md \simeq 3.9\times{10}^6$\,M$_\odot$. The median final metallicity of the sample in this case is $0.5$\,Z$_\odot$, lower than observed metallicities of some SMGs (see Section~\ref{sec:Inflows}). The metallicity and the dust mass are decreased compared to the original model because the same mass of stars enriches a larger mass of gas. The stellar mass is unchanged as this is set by the SFH, resulting in a final average gas fraction of 0.75.  Although this is larger than the average observed values in SMGs, some high redshift systems have been found with $f_g \sim 0.8$ \citep{Tacconi10,Riechers11} which is consistent with this model.  In summary, even with realistic SFHs including bursts and a larger SMG sample, we can definitively rule out a model with LIMS stardust alone, since this cannot explain the observed dust mass or the dust-to-metals ratio of the SMG population.

\subsection{Dust production in stars - adding supernovae}
\begin{figure}
\centering
\begin{minipage}{0.48\textwidth}
\includegraphics[width=8.5cm]{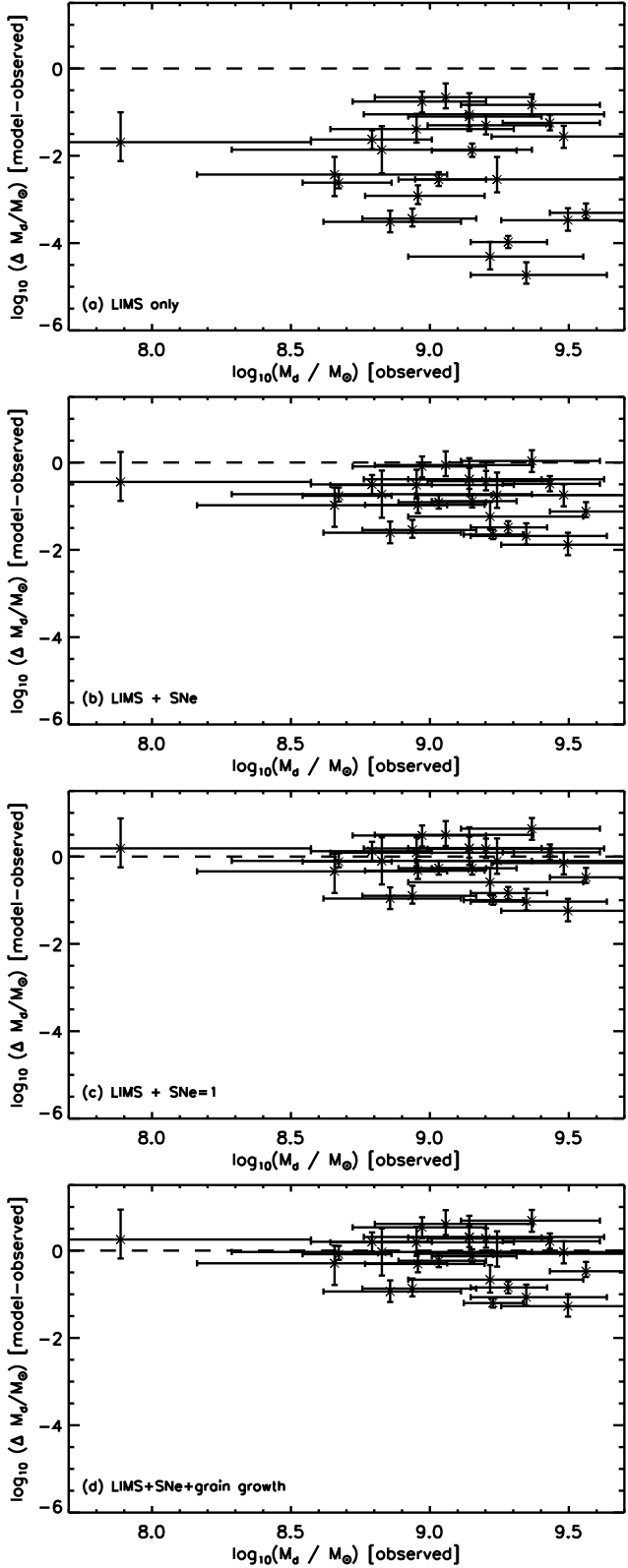}
\end{minipage}
\caption{The difference between the dust masses derived from the chemical evolution models and the observed dust masses for the high redshift SMGs, assuming no dust destruction. Dust is produced in panel (a) by low--intermediate mass stars (LIMS) only, LIMS and supernovae (b), LIMS and maximal supernova dust contribution (all metals into dust) (c), and by LIMS, SNe and grain growth (d).}
\label{fig:SFR_Mdust_Magnelli}
\end{figure}

If we include dust production from both SNe (using the yields from TF01) and LIMS, dust builds up more rapidly in SMGs with a delay of only tens of Myrs between the highest mass stars forming and evolving to the SN phase. This is evident in Fig.~\ref{fig:chem_ev_Mstar_Mdust}, as the dust produced by both SNe and LIMS closely tracks the stellar mass build-up over time, with bursts of star formation resulting in an almost instantaneous increase in the dust mass. Adding dust from SNe accounts for more than an order of magnitude increase in the dust mass of SMGs (with a median mass of $1.9\times{10}^8$\,M$_\odot$ for the LIMS+SNe model) compared to the dust mass from LIMS only ($5.0\times{10}^6$\,M$_\odot$). The model dust masses using LIMS and SNe match the observed values (accounting for the $\pm0.2$\,dex uncertainty in the {\sc magphys}-derived dust masses) in $\sim 19$\,per\,cent of cases (Table~\ref{tab:mdust_summary}). The median metallicity of the SMGs in this model is the same as with LIMS only ($0.9$\,Z$_\odot$), but with the inclusion of SN dust the median fraction of metals in the ISM in the form of dust is higher ($\eta_Z = 16$\,per cent). In Fig.~\ref{fig:SFR_Mdust_Magnelli}b it can be seen that the predicted dust masses for the majority of the SMGs falls short of the observed dust masses, which indicates additional sources of dust, or even higher SN dust yields than TF01 (where $\delta_{\rm sn} \sim 0.3-0.8$, Table~\ref{sn_yields}) are required.

In Fig~\ref{fig:SFR_Mdust_Magnelli}~(c) we consider the extreme case of maximal dust production from SN e.g. $\delta_{\rm sn} = 1$, such that all metals ejected in this phase are incorporated into dust. Sufficiently high dust masses are achieved in this scenario (a median of $7.6\times{10}^8$\,M$_\odot$) to account for the observed dust in 15/26 SMGs (see the black dashed lines in Fig.~\ref{fig:chem_ev_Mstar_Mdust}). The gas-to-dust ratio for the maximal SN dust model is in agreement with observed values for SMGs of $28^{+14}_{-11}$ and $42\pm25$\footnote{These values have been corrected for the different $\kappa_{850}$ value used compared to {\sc magphys} i.e. $(0.15/0.077)$. When we compare our model values of the total gas-to-dust ratio to observations this assumes that the observed molecular gas component dominates over the atomic gas in SMGs.} \citep{Kovacs06, Swinbank14}. The resulting median fraction of metals in the ISM in the form of dust is $\eta_Z \sim 68$\,per cent, somewhat higher than the values observed in nearby galaxies (\citealt{James02}; \citealt[and references therein]{Watson11}) and in galaxies out to $z\sim6$ \citep{Zafar13}. {\it Herschel} observations of SNRs do indicate high condensation efficiencies with $\delta_{\rm sn} \sim 0.1-1.0$ (see e.g. Table~\ref{sn_yields}), suggesting that at some point, the majority of heavy elements produced in SN ejecta can condense into dust.  However, evidence for efficient dust production is observed in only a handful of SNRs, and it is unclear how much dust will survive passage into the ISM. Furthermore uncertainties in the progenitor mass makes $\delta_{\rm sn}$ difficult to estimate observationally. It is therefore still possible that SN dust is not able to account for all the dust in galaxies, i.e. either a significant mass of dust must form rapidly in the ISM (see Section~\ref{Grain_growth}) or theoretical metal yields from SNe obtained from stellar evolution models are \emph{systematically underestimated}. Whilst differences between SFHs can cause variations in dust mass, as explored in Section~\ref{realisticsfh} this is only a factor of 1.5 for models with LIMS and SN dust. Since this is much smaller than the dust shortfall our conclusions are robust to differences between SFHs.

\subsection{The Effects of dust destruction}
\label{Dust_destruction}
There are large uncertainties about the effectiveness of dust destruction in the ISM, yet it is often used in chemical evolution analyses.  If SN shocks are efficient in destroying dust, then assuming $1000\,$M$_\odot$ of ISM is cleared by each supernova event, the dust mass we obtain is reduced by a factor of 6--10 on average if dust is produced by LIMS and LIMS+SNe, respectively. This compounds the dust budget crisis further. The effect of dust destruction on the build-up of dust is shown as the dark blue line in Fig.~\ref{fig:chem_ev_Mstar_Mdust}. With LIMS as the only source of dust, the dust destruction efficiency in our model is similar to the maximum dust destruction case in \citet{Gall11a} with $\mISM=800\,$M$_\odot$. If dust is produced by both LIMS and SNe, then any increase in the dust mass by including SNe is effectively cancelled out by the dust destroyed, resulting in a median dust mass of $1.8\times{10}^7\,$M$_\odot$, approximately 3.6 times the mass obtained with the LIMS only model ($5\times10^6$\,M$_\odot$).

If the dust is shielded in cold, dense regions of the ISM, then it is possible that the dust destruction efficiency of SN shocks will be reduced. Lower dust destruction rates have been suggested by \citet{Dwek07,Dwek11} and \citet{Gall11a}, who struggle to produce the dust masses of high redshift galaxies with efficient dust destruction.  These authors suggest that $\mISM=100\,$M$_\odot$ may be more appropriate given the increased density of the ISM gas ($n>10^3\,\rm cm^{-3}$, e.g. \citealp{Alaghband13}, see Fig.~2 in \citealp{Dwek07}) compared to the Milky Way. Including dust destruction in the model with $\mISM=100\,$M$_\odot$ now lowers the dust mass by a factor of 1.2--1.6 (compared to the LIMS only and LIMS+SNe models). A cautionary note here is that theoretical dust destruction models are not well understood or appear to be {\it too} efficient at destroying dust grains, the very fact that we observe so much dust in galaxies, including those with very little recent star formation \citep[e.g.][]{Rowlands12}, implies the destruction rate must be balanced by the injection rate from stars and another source of dust e.g. grain growth in the ISM \citep[see also][]{MWH10,Dunne11,Mattsson_Andersen12b,Boyer12}.

\begin{table*}
\begin{center}
\begin{tabular}{ >{\raggedright\arraybackslash}m{3.7cm} >{\centering\arraybackslash}m{1.95cm} >{\centering\arraybackslash}m{2.0cm}> {\centering\arraybackslash}m{2.7cm} >{\centering\arraybackslash}m{2.5cm} >{\centering\arraybackslash}m{2.2cm}}
\hline
Model & $f_{\rm{gas}}$ & $Z/$Z$_\odot$ & log$_{10}(M_{\rm{\rm dust}}/M_\ast$) & $\eta_Z$ & $\eta_g$ \\
\hline
\vspace{0.1cm}
I. Maximum stellar sources &  0.48\mbox{} {\bf 0.49} \mbox{} 0.50&  0.81\mbox{} {\bf  0.90} \mbox{}  1.06& $-$1.99\mbox{} {\bf $-1.95$} \mbox{} $-$1.90& 0.663\mbox{} {\bf 0.679} \mbox{} 0.686&    75\mbox{} {\bf    88} \mbox{}    99\\
($\delta_{\rm lims} = 0.45$, $\delta_{\rm sn}=1$) & & & & & \\
\vspace{0.3cm}
II. Stellar and interstellar &   0.48\mbox{} {\bf 0.49} \mbox{} 0.50&  0.81\mbox{} {\bf  0.90} \mbox{}  1.06& $-$2.01\mbox{} {\bf $-$1.90} \mbox{} $-$1.80& 0.652\mbox{} {\bf 0.770} \mbox{} 0.880&    56\mbox{} {\bf    79} \mbox{}   102\\
($\delta_{\rm lims} = 0.45$, $\delta_{\rm sn}\sim 0.2$, $\epsilon=500$) & & & & & \\
\vspace{0.3cm}
III. Stellar + interstellar + destruction & 0.48\mbox{} {\bf 0.49} \mbox{} 0.50&  0.81\mbox{} {\bf  0.90} \mbox{}  1.06& $-$2.13\mbox{} {\bf $-$2.01} \mbox{} $-$1.87& 0.510\mbox{} {\bf 0.599} \mbox{} 0.751&    66\mbox{} {\bf   101} \mbox{}   133\\
($\delta_{\rm lims} = 0.45$, $\delta_{\rm sn}\sim 0.2$, $\epsilon=500$, $\mISM=100\,$M$_\odot$)  & & & & & \\
\vspace{0.3cm}
IV. Stellar + interstellar + destruction + inflow + outflow &  0.48\mbox{} {\bf 0.49} \mbox{} 0.50&  0.54\mbox{} {\bf  0.57} \mbox{}  0.62& $-$2.45\mbox{} {\bf $-$2.36} \mbox{} $-$2.25& 0.360\mbox{} {\bf 0.407} \mbox{} 0.518&   166\mbox{} {\bf   231} \mbox{}   281\\
($\delta_{\rm lims} = 0.45$, $\delta_{\rm sn}\sim 0.2$, $\epsilon=500$, $\mISM=100\,$M$_\odot$, $I=1\times$\,SFR, $O=1\times$\,SFR)  & & & & & \\
\hline
\end{tabular}     
\caption[Summary of the SMG properties derived from different chemical evolution models.]{Summary of the properties derived from different chemical evolution models which best describe most of the observed average properties of the 26 $z>1$ SMGs (see Table~\ref{tab:chem_ev_summary_all} for a list of all the model results).  The properties are: the final gas fraction $f_{\rm gas}$, metallicity in units of solar metallicity ($Z$; the ratio of metal mass to gas mass, with Z$_\odot=0.019$), the dust-to-stellar mass ratio (\mdms), the dust-to-metal mass ratio ($\eta_Z = M_d/M_Z$) and the gas-to-dust ratio ($\eta_g = M_g/M_d$). For reference, the average $f_g$ of SMGs is $30-50$\,per\,cent \citep{Tacconi08,Riechers11}; the typical $Z$ is Solar or slightly subsolar \citep{Swinbank04,Banerji12,Nagao12}; the mean log$_{10}$(\mdms) is $-$1.71; the typical $\eta_Z$ is $\sim0.5$ \citep{Zafar13}; and average $\eta_g$ values are $\sim30-50$ \citep{Kovacs06, Swinbank14}. For each chemical evolution model we list the median value of the sample in bold and the 16th and 84th percentiles to indicate the range of values in the sample.}
\label{tab:chem_ev_summary}
\end{center}
\end{table*}

\subsection{Adding grain growth}
\label{Grain_growth}

In order to obtain a minimum $\tau_{\rm grow} \sim (20-200)$\,Myr in our fiducial models, in line with expected grain growth timescales and to relieve the `budget crisis' (\citealp{Zhukovska08}, \citealp*{Mattsson_Andersen12a}), we set $\epsilon=500$ (Section~\ref{Grain_growth_intro}, Eq.~\ref{eq:graingrowth}). If the value of $\epsilon$ is lower, then the grain growth timescale is longer which reduces the dust mass produced by grain growth at a given age. It is possible that $\epsilon$ is larger than the value adopted here, however, all of the metals in each SMG are rapidly incorporated into dust grains.

The effect of adding grain growth to the LIMS only model (with no dust destruction) on the dust mass is shown in Fig.~\ref{fig:chem_ev_Mstar_Mdust} by the solid red line. However, this model fails to adequately reproduce the observed dust masses for the majority of SMGs. In Fig~\ref{fig:SFR_Mdust_Magnelli}d we therefore consider dust produced by LIMS, SNe (using TF01 yields) and grain growth in the ISM. We find that including grain growth on average increases the dust mass by a factor of 200 to $\simeq 9.8\times{10}^8\,$M$_\odot$, compared to a model in which dust is contributed by LIMS only. Grain growth can therefore easily make up the shortfall in the predicted dust masses for 60-70\,per\,cent of SMGs in the sample (Table~\ref{tab:mdust_summary}), but only if {\it grain growth is the dominant form of dust production in SMGs} (see also \citealt{Mattsson2011}). Indeed, this model can easily account for the observed dust masses of the majority of SMGs in our sample, however, in this scenario, a large fraction (77\,per\,cent) of the metals is locked up in the form of dust which is higher than typically observed. At first glance the large fraction of metals locked up in dust is in conflict with \citet{Zafar13} who found a relatively constant dust-to-metals ratio of 0.5 out to $z\sim6$. However, the uncertainties on the dust and metal masses are likely to be a factor of a few, therefore the average SMG dust-to-metals ratio is not strictly in disagreement with the results of \citet{Zafar13}.

In order to explain all of the dust in every SMG in our sample we run the most optimistic models where dust is produced by LIMS, SNe (with all metals incorporated into dust) and grain growth, with no destruction. Even with such high dust condensation efficiencies we cannot account for the dust in all SMGs in our sample, only reproducing the dust masses for $\sim 69$\,per\,cent of SMGs.  Whilst the average dust mass and gas-to-dust ratio in this scenario agree well with observed values, around 96\,per\,cent of metals are in the form of dust, which is much higher than observed. Given that the majority of metals are already in dust grains, increasing the grain growth efficiency does not substantially improve the dust yield and therefore grain growth {\it cannot} explain all of this shortfall. The close agreement in the average dust mass for the SMG sample ($1.2^{+0.3}_{-0.2}\times{10}^9\,$M$_\odot$) and the average mass of metals (assuming a metallicity of $\sim Z_\odot$ and a gas mass of $5\times{10}^{10}\,$M$_\odot$) further suggests that metal yields of stars may be systematically underestimated (see also \citealt{Mattsson2011}).

Ultimately, if efficient dust destruction is included along with LIMS, SN dust and grain growth, then the dust produced in these galaxies is not enough to account for the observed dust masses, with the median dust mass reaching $6.0\times{10}^7\,$M$_\odot$. For the SMGs whose predicted dust masses fall short of the observed value, it is possible that dust destruction is less efficient than assumed in the literature. Whilst there are considerable uncertainties in the sources of dust production and destruction in galaxies, we can definitively state that LIMS cannot be the only source of dust in SMGs, and show that this result is robust to larger samples and bursty SFHs. In order to explain the observed dust masses of 19\,per\,cent of the SMGs in this work, we require dust from LIMS and $\sim$20\,per\,cent of metals produced in SN to condense into dust grains \emph{and survive} ($\delta_{\rm dest}=0$). If (as is more likely) $\delta_{\rm dest}>0$, another dust source is required to account for the observed masses. In the next sections, we look at the effects of inflows and outflows on the dust properties of SMGs.

\subsection{Inflows}
\label{sec:Inflows}
The closed box chemical evolution model for the SMGS produces a (median) final metallicity of $0.9$\,Z$_\odot$ (Table~\ref{tab:chem_ev_summary} \& Table~\ref{tab:chem_ev_summary_all}) in line with Solar/sub-Solar metallicities observed in \citet{Swinbank04} and \citet{Banerji12}.  However, although the closed box model is the simplest approach to chemical evolution, in reality, galaxies are unlikely to be closed systems (e.g. \citealt{Erb08}).  Examples of this include the well known G-Dwarf problem, which requires infall of material in the Milky Way \citep[e.g.][]{vandenBergh62, SearleSargent72, PagelPatchett75, Tinsley1980}, and the observed wide range in stellar metallicity for galaxies with fixed gas-phase oxygen abundances \citep{Gallazzi05} which also requires inflows and outflows of gas.  As yet, there is limited direct observational evidence for gas inflows, but recent studies at high redshift provide further indirect evidence that inflows are required, including the need to sustain high SFRs \citep{Giavalisco11, Reddy12b, Tacconi13} as well as being an essential ingredient in galaxy formation simulations at these epochs \citep[e.g.][]{Dekel09}.  Due to the importance of gas accretion in galaxy evolution models, we therefore investigate the effect of inflows on our results from the chemical evolution model. In general, inflows will act to decrease the gas-phase metallicity (as the enriched interstellar gas is diluted by the metal-poor inflow, \citealt{Edmunds90, Edmunds_Eales98, Edmunds01}), though the dust mass contributed by stellar sources is largely unchanged. Conversely, inflows will have a more significant effect on the dust produced by grain growth by decreasing the grain growth timescale (Eq.~\ref{eq:graingrowth}).  To determine the `inflow prescription' to include in this work, we use an inflow rate on the order of the SFR to be consistent with the semi-analytic model of \citet{Dutton10} and \citet{Erb08}, who find that the rate-of-change of the gas mass (inflow-outflow) is in a steady state with the SFR.

We initially assume that an inflow delivers metal-free gas ($Z_I=0$, Eq.~\ref{eq:metals} and $(M_d/M_g)_I=0$, Eq.~\ref{eq:chemdust}) to the galaxy at a rate proportional to the SFR throughout the lifetime of each SMG. We tune the initial gas mass such that the SMGs have the same final gas fraction as the closed box model $\sim0.5$ to provide a consistent comparison. To demonstrate the effect of inflow on the chemical evolution of SMGs we run the model including dust from LIMS only, with no destruction or grain growth. We find that an inflow rate equal to the SFR reduces the median metallicity of the SMG sample to $0.7$\,Z$_\odot$, whilst the median dust mass is increased by a factor of 1.3 (Table~\ref{tab:chem_ev_summary} \& Table~\ref{tab:chem_ev_summary_all}). This is because the inflow model starts with a smaller mass of gas which is enriched to a higher metallicity than the closed box model, which results in a higher dust mass. As the inflow of pristine gas continues, this dilutes the metallicity of the gas but does not affect the mass of dust in the galaxy.

The inflow rates we require to match the observed properties of SMGs are $<1\times$SFR (for the SMG sample the 16th--84th percentile range of the median SFR over time is 6--600\,M$_\odot$yr$^{-1}$), which is consistent with indirect observational support from some studies of high redshift galaxies, but higher than the gas accretion rates required in simulations (typically 40-60\,M$_\odot$yr$^{-1}$, \citealt{Keres05,Dave10, vandeVoort11a}). Unfortunately, the large range in the observed metallicities of SMGs and uncertainties due to possible AGN contamination of the emission lines do not allow us to discriminate between models which have different gas inflow rates. 

\subsection{Outflows}
\label{sec:Outflows}
Outflows of gas are thought to be common in actively star-forming galaxies at all epochs \citep{Heckman00, Weiner09, Rubin10, Diamond-Stanic12, Bradshaw13}, and may be either driven by stars (stellar winds and SN), or by AGN.   Significant outflows of material are implied by the results of \citet{Menard10}, who found evidence for dust in galaxy halos with a mass comparable to that of dust in the disk.  Furthermore, \citet{Erb06} suggest that the mass--metallicity relation at $z\sim2$ is modulated by metal-rich outflows from galaxies, with rates of up to four times the SFR. The upper end of their range agrees with results from \citet{Dunne11} who used a simple analytic chemical evolution model to demonstrate that outflow rates of four times the SFR best describes the evolution of the dust mass function of H-ATLAS galaxies at $0<z<0.5$. Outflows could therefore be responsible for the significant metal enrichment of the IGM  and are an important component of the chemical evolution of galaxies. 

In this work, we assume that the gas and dust in the ISM are well mixed  so that $Z_O=Z$ (so-called unenriched outflow, \citealt{Dalcanton07}, see also Eq.~\ref{eq:metals}), and that outflows remove material (including gas, metals and dust) from the galaxy at a rate proportional to the SFR. To demonstrate the effect of outflows on the chemical evolution of SMGs we again run the model including only dust from LIMS, with no destruction or grain growth. In general, outflows decrease the overall dust mass (though not the contribution from stellar sources) and decrease the ISM metallicity.  In this work, as with the inflow model, the initial gas mass is tuned such that the SMGs have a final gas fraction of $0.5$. Outflows of $1\times$ the SFR reduce the dust mass in SMGs on average by a factor of 1.4, and the ISM metallicity is reduced to $0.7$\,Z$_\odot$. This is well within the large range of the observed dust masses and metallicities of SMGs, therefore outflows of gas equal to the SFR can be accommodated in our chemical evolution models.

It is also possible that both inflows and outflows occur simultaneously \citep{Sakamoto13}, or in short succession \citep{Dalcanton07}. By allowing simultaneous inflow and outflow in our model, with rates equal to the SFR, the metallicity is decreased to $0.6$\,Z$_\odot$.
The dust mass from LIMS is reduced by a factor of 1.3 compared to the closed box model, which is well within the observational range of SMG dust masses. The amount of dust removed from the galaxy in the outflow model is much lower than than that suggested by the \citet{Menard10} results, which imply a higher outflow rate is needed to remove half of the dust mass. However, outflow rates significantly larger than the rate of gas inflow would serve to decrease the dust mass, which increases the tension between model and observed dust masses and compounds the dust budget crisis further.

\subsection{Variations in the IMF}
The amount of dust formed in galaxies is strongly linked to the amount of metals produced by stars.  Increasing the yields and decreasing the amount of low mass `dead' stars produced in a stellar population by varying the IMF is therefore one way to possibly solve the dust budget crisis.  Previous works \citep[e.g.][]{Dwek11,Gall11Rev,Valiante11} have shown that by invoking a top-heavy IMF, one can easily reproduce the observed dust masses in some high redshift galaxies, but it is unclear whether the IMF in other galaxies is Milky-Way like and invariant with time and location (see the review in \citealp*{Bastian10}). Many studies have suggested that a top-heavy IMF is a natural consequence of the extreme environment in high-redshift galaxies, for example due to bursty SFHs and the denser ISM in comparison to local galaxies (\citealp*{Dabringhausen09}; \citealp{Papadopoulos11, Kroupa12}). Indeed \citet{Gunawardhana11} found evidence for a strong relationship between SFR and IMF slope for $z<0.3$ galaxies, such that galaxies with higher SFRs form more massive stars in a given stellar population. Here we investigate the sensitivity of the derived dust mass to the IMF in the models and whether this allows us to predict the slope of the IMF required to resolve the dust budget crisis in SMGs.

We increase the power law slope of the Chabrier IMF ($\phi(m) \propto m^{-\alpha}$ where $\alpha$ is the slope) from $-1.3$ to $-0.67$, but leave the low mass end ($<1$M$_{\odot}$) unchanged. The value of the high mass slope is found by extrapolating the relationship of \citet{Gunawardhana11} between IMF slope and SFR for low redshift star-forming galaxies to the average SMG SFR ($390\,$M$_\odot$yr$^{-1}$). Using the fiducial SFHs introduced in Section~\ref{sec:chem_ev_description} i.e. an exponentially declining SFH with $\psi(0) = 150$\,M$_\odot$yr$^{-1}$, an exponentially increasing SFH with $\psi(f) = 150$\,M$_\odot$yr$^{-1}$, a constant SFR of 150\,M$_\odot$yr$^{-1}$, and an instantaneous burst, a \emph{top-heavy IMF does not increase the dust mass enough to account for the observed dust masses of SMGs with an LIMS-source of dust only}, even with the increase in the number of `super-AGBs' formed. Considering dust production from both LIMS and SNe and using the fiducial SFHs, we find that an IMF slope of $-0.67$ reproduces the average observed SMG dust masses ($1.2^{+0.3}_{-0.2}\times{10}^9\,$M$_\odot$) within a factor of two. This slight shortfall in dust mass can be alleviated by including a small amount of grain growth which allows the average SMG dust mass to be easily reached.

A top-heavy IMF also leads to a higher destruction rate because of the increased SN rate (Eqs.~\ref{eq:Dust_destruction} \& \ref{eq:snrate}, \citealt{Gall11a, Mattsson2011}). Therefore the increase in the dust mass from LIMS and SNe with efficient destruction ($\mISM=1000$\,M$_\odot$) achieved with a top-heavy IMF is on average a factor of 1.3 lower compared to a Chabrier IMF with the same dust sources and destruction rates (at a time of 0.5\,Gyr after the onset of star formation). In summary, invoking a top-heavy IMF with no dust destruction and no grain growth can solve the dust budget crisis in SMGs, but given the uncertainties involved, the high dust masses do not provide unequivocal evidence for a top-heavy IMF.

\subsection{The Dust Emissivity Caveat}
\label{sec:kappa}
On a final note, it is also possible that dust produced by SNe in high redshift galaxies has different properties to that assumed in this work, where the dust emissivity $\kappa$ used to determine the dust mass is calibrated from observations of the Milky Way and other nearby galaxies (Section~\ref{sec:sample_selection}, see also \citealp{Valiante11}).  If, for example, the dust emissivity in high redshift SMGs was systematically higher due to changes in the dust grain size distribution, shape or grain composition, this would serve to decrease the observed dust masses thereby alleviating the tension between observed and predicted properties of SMGs. One possibility to increase $\kappa$ is if the dust grains are in dense gas environments and have amorphous structures \citep{Ossenkopf_Henning94}. Although there is no observational evidence for different grain properties in SMGs, recent FIR observations of the Magellanic Clouds found that dust properties may be different from those in the Milky Way \citep{Meixner10} due to a recent increase in the Type II SN rate. It is therefore plausible that the dust grain properties may be different in environments where dust is produced and/or reprocessed by supernovae. Indeed \citet{Bianchi07} derive $\kappa$ values for freshly-formed SN dust and SN-processed dust, with the latter case predicting a $\kappa_{850}$ value\footnote{Scaled using $\beta=1.4$ as appropriate for their processed grain model.} which is 2.46 times higher than the value assumed in this work. If dust is produced by LIMS only, $\kappa$ would need to differ by a factor of 240 compared to the MW in order to resolve the dust budget crisis in this sample. Adding a SNe dust source with no grain growth or dust destruction, the crisis can be solved with $\kappa$ increasing on average by a factor of 7 at $1<z<5$. Note that to explain the dust shortfall in individual SMGs by dust emissivity variation alone, $\kappa_{850}$ would need to vary from $0.074-5.9\,\rm{m}^2\rm{kg}^{-1}$. If we include efficient dust destruction in our models then $\kappa_{850}$ would need to be a factor of 70 higher (i.e. $5.4\,\rm{m}^2\rm{kg}^{-1}$) at $1<z<5$ to solve the dust budget crisis. In comparison, the range of $\kappa_{850}$ in the literature \citep[][see references therein]{Valiante11} is $0.02-2.0\,\rm{m}^2\rm{kg}^{-1}$ where the higher end of this range refers to the results from theoretical simulations of dust processed by SN shocks \citep{Bianchi07}. Although variations in the dust emissivity by up to a factor of 10 are theoretically possible \citep{Ossenkopf_Henning94}, there is currently no observational evidence to suggest the large variations in the dust emissivity which would be required to solve the crisis in this sample. Furthermore, \citet{Rowlands14b} suggest that due to the consistency between observations of gas and dust in individual SMGs and the gas-to-dust ratios implied by the ratio of FIR to CO luminosity, $\kappa$ does not evolve strongly between low redshift dusty galaxies and SMGs.

\subsection{Reproducing SMG properties}
\label{sec:Sensible_models}
In summary, the models which best reproduce the dust-to-stellar mass and gas-to-dust properties of SMGs are LIMS+SNe+grain growth (62\,per\,cent of the sample) and LIMS+maximal SN dust production (explains 58\,per\,cent of dust masses in the SMGs sample, although this is unphysical as it requires all of the SN metal yields to be in the form of dust). The most plausible models are summarised in Table~\ref{tab:chem_ev_summary}. In reality a mixture of these models will most likely best describe the properties of high redshift SMGs.

Based on these findings, more complex models were run with dust created by LIMS, SNe and grain growth with moderate dust destruction ($\mISM=100\,$M$_\odot$) to reduce the dust-to-metals ratio slightly to better match observations of SMGs. For a closed box model the average dust-to-stellar mass and gas-to-dust ratios agree well with observed values for SMGs (see Table~\ref{tab:chem_ev_summary}). This shows that a modest amount of dust destruction can be accommodated if dust is produced by both \emph{stellar and interstellar sources}. In reality galaxies are unlikely to be closed boxes, we therefore run the same models but with an inflow and outflow with a rate equal to the SFR. Compared to the closed box these models predict slightly lower average final \mdms (by a factor of two) and slightly higher gas-to-dust ratios (by a factor of two), but these differences are still well within the observational range of these parameters for SMGs.

\section{Conclusions}
\label{sec:conclusions}

In this paper we have used an updated chemical evolution model to reproduce the properties of a submillimetre selected sample of 26 massive, dusty galaxies in the redshift range $1.0<z<5.3$. Our chemical modelling for the first time utilises complex SFHs derived from SED fitting to the the UV--submillimetre photometry and a detailed treatment of the dust sources and sinks in galaxies.  

We can rule out a number of models (Table~\ref{tab:chem_ev_summary_all}) which result in dust-to-stellar masses and/or gas-to-dust ratios which are inconsistent with observations of SMGs. These models include those with dust produced by LIMS only, and those which have efficient dust destruction (mass of ISM $\mISM=1000\,$M$_\odot$ cleared of dust). The models which best match the observed gas-to-dust ratios include rapid dust build-up from grain growth and supernova dust sources. Our main results are as follows:

\begin{itemize}

\item We find that dust produced only by low--intermediate mass stars (LIMS) falls a factor 240 short of the observed dust masses of SMGs. Adding an extra source of dust from supernovae can account for the dust mass in SMGs in only 19\,per\,cent\, of cases. Even after accounting for dust produced by supernovae, the remaining deficit in the dust mass budget suggests that higher supernova metal yields, and/or substantial grain growth are required in order for the dust mass predicted by the chemical evolution models to match observations of SMGs.

\item Efficient destruction of dust grains by supernova shocks ($\mISM=1000\,$M$_\odot$) on average decreases the dust mass from LIMS+SNe by a factor of 6--10. Additional sources of dust are required in order to account for the additional shortfall of dust in SMGs caused by dust destruction. Alternatively, dust destruction may be less efficient if dust grains are shielded from supernova shocks in dense regions of the ISM. A small amount of dust destruction ($\mISM=100\,$M$_\odot$) can be accommodated in our models only if dust is produced efficiently by \emph{both} stellar and interstellar sources.

\item The average metallicity in the closed box model reaches $0.9$\,Z$_\odot$, which is consistent with the metallicity measured in SMGs. If inflows of pristine gas occur with a rate equal to the SFR the metallicity is reduced to $0.7$\,Z$_\odot$; a similar metallicity is reached with enriched gas outflows. Inflows and outflows result in a modest decrease of a factor $<1.5$ in the dust mass of SMGs. Given the current large range in observed gas-phase metallicities in SMGs, and uncertainties due to possible AGN contamination of the emission lines, we cannot currently distinguish between different inflow and outflow rates.  Measurements of gas-phase metallicities which are not affected by the presence of an AGN are required for larger samples of SMGs.

\item A top-heavy IMF cannot account for the observed dust masses if dust is produced by LIMS only. With no dust destruction we found that a top-heavy IMF with dust produced by both LIMS and supernovae can produce the average dust mass observed in SMGs (within a factor of two) therefore resolving the dust budget crisis. Yet, given the uncertainties involved (e.g. in the dust destruction rate and metallicity in SMGs) this does not provide unequivocal evidence for a top-heavy IMF in dusty high redshift galaxies.

\item Increasing the dust emissivity on average by a factor of 7 can solve the dust budget crisis if dust is produced by LIMS and SNe and is not destroyed by supernova shocks. Variations in the dust emissivity are theoretically predicted to be a factor of $<3$, and, currently there is no observational evidence to suggest a large variation in emissivity occurs in high-redshift SMGs. Finally, an alternative explanation for the dust budget crisis is that the metal yields of stars may be systematically underestimated.

\end{itemize}

\section*{Acknowledgements}
We sincerely thank the referee, Raffaella Schneider, for her useful suggestions which have much improved the clarity of this manuscript. We thank David Elbaz, Fraser Pearce, Ian Smail, Michal Micha{\l}owski and Lars Mattsson for helpful comments. K.~R. acknowledges support from the European Research Council Starting Grant (P.I. V.~Wild). H.~L.~G acknowledges support from the Science, Technology and Facilities Council. This research has made use of data from the HerMES project (http://hermes.sussex.ac.uk/). HerMES is a Herschel Key Programme utilising Guaranteed Time from the SPIRE instrument team,  ESAC scientists and a mission scientist. HerMES is described in \citet{Oliver10_Hermes}.

\bibliographystyle{mn2e}
\bibliography{references}

\appendix

\section{Results of chemical evolution models}

\begin{table*}
\begin{center}
\caption[Summary of the properties derived from different chemical evolution models.]{Summary of the properties derived from different chemical evolution models for the 26 SMGs, which have a mean dust mass of $1.2^{+0.3}_{-0.2}\times{10}^9\,$M$_\odot$. The properties are: the final gas fraction $f_{\rm gas}$, metallicity in units of solar metallicity ($Z$; the ratio of metal mass to gas mass, with Z$_\odot=0.019$), the dust-to-stellar mass ratio (\mdms), the dust-to-metal mass ratio ($\eta_Z = M_{\rm{\rm dust}}/M_Z$) and the gas-to-dust ratio ($\eta_g = M_{\rm{gas}}/M_{\rm{\rm dust}}$). For reference, the average $f_{\rm gas}$ of SMGs is $30-50$\,per\,cent \citep{Tacconi08,Riechers11}; the typical $Z$ is Solar or sub-Solar \citep[as low as $\sim0.2$Z$_\odot$;][]{Swinbank04,Banerji12,Nagao12}; the mean log$_{10}$(\mdms) is $-$1.71; the typical $\eta_Z$ is $\sim0.5$ \citep{Zafar13}; and average $\eta_g$ values are $\sim30-50$ \citep{Kovacs06, Swinbank14}. For each chemical evolution model we list the median value in bold and the 16th and 84th percentiles to indicate the spread of values in the sample. A tick (cross) indicates that the model does (does not) provide a plausible match to observations of SMGs.}
\begin{tabular}{ >{\raggedright\arraybackslash}m{3.5cm} >{\centering\arraybackslash}m{1.95cm} >{\centering\arraybackslash}m{2.0cm}> {\centering\arraybackslash}m{2.7cm} >{\centering\arraybackslash}m{2.5cm} >{\centering\arraybackslash}m{2.8cm}}
\hline
Model & $f_{\rm{gas}}$ & $Z/$Z$_\odot$ & log$_{10}(M_{\rm{\rm dust}}/M_\ast$) & $\eta_Z$ & $\eta_g$ \\
\hline
\vspace{0.1cm}
\cross LIMS dust only &  0.48\mbox{} {\bf 0.49} \mbox{} 0.50&  0.81\mbox{} {\bf  0.90} \mbox{}  1.06& -4.81\mbox{} {\bf -3.81} \mbox{} -3.23& 0.001\mbox{} {\bf 0.009} \mbox{} 0.031&  1600\mbox{} {\bf 10900} \mbox{} 66400\\
\vspace{0.3cm}
\cross LIMS+supernova dust &  0.48\mbox{} {\bf 0.49} \mbox{} 0.50&  0.81\mbox{} {\bf  0.90} \mbox{}  1.06& -2.65\mbox{} {\bf -2.56} \mbox{} -2.50& 0.150\mbox{} {\bf 0.157} \mbox{} 0.176&   297\mbox{} {\bf   384} \mbox{}   445\\
\vspace{0.3cm}
\tick LIMS+maximal supernova dust &  0.48\mbox{} {\bf 0.49} \mbox{} 0.50&  0.81\mbox{} {\bf  0.90} \mbox{}  1.06& -1.99\mbox{} {\bf -1.95} \mbox{} -1.90& 0.663\mbox{} {\bf 0.679} \mbox{} 0.686&    75\mbox{} {\bf    88} \mbox{}    99\\
\vspace{0.3cm}
\cross LIMS only+destruction ($\mISM=1000\,$M$_\odot$) &  0.48\mbox{} {\bf 0.49} \mbox{} 0.50&  0.81\mbox{} {\bf  0.90} \mbox{}  1.06& -5.53\mbox{} {\bf -4.72} \mbox{} -4.06& 0.000\mbox{} {\bf 0.001} \mbox{} 0.005& 10700\mbox{} {\bf 55100} \mbox{}335000\\
\vspace{0.3cm}
\cross LIMS only+destruction ($\mISM=100\,$M$_\odot$) &  0.48\mbox{} {\bf 0.49} \mbox{} 0.50&  0.81\mbox{} {\bf  0.90} \mbox{}  1.06& -4.88\mbox{} {\bf -3.92} \mbox{} -3.39& 0.001\mbox{} {\bf 0.007} \mbox{} 0.021&  2310\mbox{} {\bf 15000} \mbox{} 76400\\
\vspace{0.3cm}
\cross LIMS+supernova dust+destruction ($\mISM=1000\,$M$_\odot$) &  0.48\mbox{} {\bf 0.49} \mbox{} 0.50&  0.81\mbox{} {\bf  0.90} \mbox{}  1.06& -3.55\mbox{} {\bf -3.53} \mbox{} -3.44& 0.018\mbox{} {\bf 0.018} \mbox{} 0.022&  2560\mbox{} {\bf  3320} \mbox{}  3490\\
\vspace{0.3cm}
\cross LIMS + maximal supernova dust + grain growth&  0.48\mbox{} {\bf 0.49} \mbox{} 0.50&  0.81\mbox{} {\bf  0.90} \mbox{}  1.06& -1.84\mbox{} {\bf -1.80} \mbox{} -1.75& 0.949\mbox{} {\bf 0.958} \mbox{} 0.964&    51\mbox{} {\bf    62} \mbox{}    69\\
\vspace{0.3cm}
\cross LIMS+supernova dust+destruction ($\mISM=100\,$M$_\odot$) &  0.48\mbox{} {\bf 0.49} \mbox{} 0.50&  0.81\mbox{} {\bf  0.90} \mbox{}  1.06& -2.80\mbox{} {\bf -2.76} \mbox{} -2.70& 0.103\mbox{} {\bf 0.105} \mbox{} 0.110&   467\mbox{} {\bf   579} \mbox{}   627\\
\vspace{0.3cm}
\cross LIMS+grain growth &  0.48\mbox{} {\bf 0.49} \mbox{} 0.50&  0.81\mbox{} {\bf  0.90} \mbox{}  1.06& -4.32\mbox{} {\bf -2.73} \mbox{} -1.97& 0.003\mbox{} {\bf 0.112} \mbox{} 0.612&    79\mbox{} {\bf   592} \mbox{} 21200\\
\vspace{0.3cm}
\cross LIMS+SNe+destruction ($\mISM=1000\,$M$_\odot$) + grain growth &  0.48\mbox{} {\bf 0.49} \mbox{} 0.50&  0.81\mbox{} {\bf  0.90} \mbox{}  1.06& -3.18\mbox{} {\bf -3.09} \mbox{} -2.91& 0.043\mbox{} {\bf 0.048} \mbox{} 0.075&   770\mbox{} {\bf  1210} \mbox{}  1530\\
\vspace{0.3cm}
\tick LIMS+supernova dust+grain growth &  0.48\mbox{} {\bf 0.49} \mbox{} 0.50&  0.81\mbox{} {\bf  0.90} \mbox{}  1.06& -2.01\mbox{} {\bf -1.90} \mbox{} -1.80& 0.652\mbox{} {\bf 0.770} \mbox{} 0.880&    56\mbox{} {\bf    79} \mbox{}   102\\
\vspace{0.3cm}
\cross LIMS+destruction ($\mISM=1000\,$M$_\odot$) +grain growth &  0.48\mbox{} {\bf 0.49} \mbox{} 0.50&  0.81\mbox{} {\bf  0.90} \mbox{}  1.06& -5.01\mbox{} {\bf -4.29} \mbox{} -3.52& 0.001\mbox{} {\bf 0.003} \mbox{} 0.017&  3100\mbox{} {\bf 25600} \mbox{}103000\\
\vspace{0.3cm}
\cross LIMS inflow ($I=1\times$\,SFR) &  0.48\mbox{} {\bf 0.50} \mbox{} 0.53&  0.65\mbox{} {\bf  0.68} \mbox{}  0.71& -4.69\mbox{} {\bf -3.75} \mbox{} -3.25& 0.002\mbox{} {\bf 0.014} \mbox{} 0.036&  2060\mbox{} {\bf 10400} \mbox{} 45000\\
\vspace{0.3cm}
\cross LIMS outflow ($I=1\times$\,SFR) &  0.47\mbox{} {\bf 0.54} \mbox{} 0.56&  0.59\mbox{} {\bf  0.68} \mbox{}  0.83& -4.97\mbox{} {\bf -3.97} \mbox{} -3.44& 0.001\mbox{} {\bf 0.007} \mbox{} 0.025&  2520\mbox{} {\bf 19900} \mbox{}122000\\
\vspace{0.3cm}
\cross LIMS inflow+outflow ($I=O=1\times$\,SFR) &  0.48\mbox{} {\bf 0.49} \mbox{} 0.50&  0.54\mbox{} {\bf  0.57} \mbox{}  0.62& -4.88\mbox{} {\bf -3.97} \mbox{} -3.48& 0.001\mbox{} {\bf 0.010} \mbox{} 0.030&  2880\mbox{} {\bf 16400} \mbox{} 77400\\
\vspace{0.3cm}
\cross LIMS dust only ($2\times$ initial gas mass)  &  0.74\mbox{} {\bf 0.75} \mbox{} 0.75&  0.44\mbox{} {\bf  0.47} \mbox{}  0.53& -4.97\mbox{} {\bf -3.89} \mbox{} -3.27& 0.000\mbox{} {\bf 0.005} \mbox{} 0.018&  5450\mbox{} {\bf 36900} \mbox{}281000\\
\vspace{0.3cm}
\tick LIMS+SNe+destruction ($\mISM=100\,$M$_\odot$) +grain growth  &  0.48\mbox{} {\bf 0.49} \mbox{} 0.50&  0.81\mbox{} {\bf  0.90} \mbox{}  1.06& -2.13\mbox{} {\bf -2.01} \mbox{} -1.87& 0.510\mbox{} {\bf 0.599} \mbox{} 0.751&    66\mbox{} {\bf   101} \mbox{}   133\\
\vspace{0.3cm}
\tick LIMS+SNe+destruction ($\mISM=100\,$M$_\odot$) +grain growth + inflow ($I=1\times$\,SFR) outflow ($O=1\times$\,SFR) &  0.48\mbox{} {\bf 0.49} \mbox{} 0.50&  0.54\mbox{} {\bf  0.57} \mbox{}  0.62& -2.45\mbox{} {\bf -2.36} \mbox{} -2.25& 0.360\mbox{} {\bf 0.407} \mbox{} 0.518&   166\mbox{} {\bf   231} \mbox{}   281\\
\hline
\end{tabular}                                                                                                      
\label{tab:chem_ev_summary_all}
\end{center}
\end{table*}

\label{lastpage}

\end{document}